\documentclass[11pt,tightenlines,eqsecnum,floats,aps,amsmath,amssymb,nofootinbib,prd,shownopacs,floatfix,notitlepage]{revtex4-1}
\usepackage{graphicx, wrapfig}
\usepackage{amssymb}
\usepackage{color}
\usepackage{mathrsfs}
\usepackage{subfigure}
\def\f{\frac}

\def\d{\textrm{d}}

\def\lp{l_{\rm Pl}}
\def\tpl{t_{\rm Pl}}
\def\mpl{m_{\rm Pl}}
\def\spl{s_{\rm Pl}}

\def\rhopl{\rho_{\rm Pl}}

\def\ns{n_{\rm s}}
\def\as{A_{\rm s}}
\def\epsilonV{\epsilon_{_{\rm V}}}
\def\deltaV{\delta_{_{\rm V}}}
\def\ks{k_*}
\def\ts{t_*}
\def\Ns{N_*}
\def\rhomax{\rho_{_{\rm max}}}
\def\phib{\phi_{_{\rm B}}}
\def\phidb{\dot{\phi}_{_{\rm B}}}
\def\ig{\includegraphics}
\def\plqc{\mathcal P_{\rm LQC}}
\def\pbd{\mathcal P_{\rm BD}}
\def\ksLQC{k_*^{\rm LQC}}

\usepackage{enumerate}

\newcommand{\be}{\nopagebreak[3]\begin{equation}}
\newcommand{\ee}{\end{equation}}
\newcommand{\bfig}{\nopagebreak[3]\begin{figure}}
\newcommand{\efig}{\end{figure}}
\newcommand{\ba}{\nopagebreak[3]\begin{eqnarray}}
\newcommand{\ea}{\end{eqnarray}}

\newcommand{\bmult}{\nopagebreak[3]\begin{multline}}
\newcommand{\emult}{\end{multline}}
\newcommand{\fref}[1]{Fig.\,\ref{#1}}
\newcommand{\eref}[1]{eq.\,(\ref{#1})}
\newcommand{\sref}[1]{Sec.\,\ref{#1}}
\newcommand{\tref}[1]{Table\,\ref{#1}}

\begin{document}
\title{Phenomenological investigation of a quantum gravity extension of inflation with 
the Starobinsky potential}
\author{B\'eatrice Bonga}
\email{bpb165@psu.edu }
\author{Brajesh Gupt}
\email{bgupt@gravity.psu.edu}

\affiliation{
Institute for Gravitation and the Cosmos \& Physics Department, The Pennsylvania State University, University Park, PA 16802 U.S.A.
}

\pacs{}
\begin{abstract}
%Recent data from the Planck mission favors inflation with the Starobinsky potential. 
We investigate the pre-inflationary dynamics of inflation with the Starobinsky 
potential, favored by recent data from the Planck mission, using techniques 
developed to study cosmological perturbations on quantum spacetimes in the 
framework of loop quantum cosmology. We find that for a large part of the initial 
data, inflation compatible with observations occurs. There exists a subset of 
this initial data that leads to quantum gravity signatures that are potentially 
observable. Interestingly, despite the different inflationary dynamics, these 
quantum gravity corrections to the powerspectra are similar to those obtained 
for inflation with a quadratic potential, including suppression of power at 
large scales. Furthermore, for super horizon modes the tensor modes show 
deviations from the standard inflationary paradigm that are unique to the 
Starobinsky potential and could be important for non-Gaussian modulation and 
tensor fossils.
\end{abstract}

\maketitle

%%%%%%%%%%%%%%%%%%%%%%%%%%%%%%%%%%%%%%%%%%%%%%%%%%%%%%%%%%%%%%%%%%
\section{Introduction}\label{sec:intro}
The paradigm of cosmic inflation is the most accepted one that explains the 
origin of the anisotropy observed in the cosmic microwave background (CMB). The 
inflationary phase of accelerated expansion stretches tiny primordial quantum 
inhomogeneities to large scale perturbations which in turn seed the large 
scale structure observed today \cite{guth,albrecht,linde,lindechaotic,
liddlelyth, mukhanov,weinberg}. This is a remarkable accomplishment of 
cosmology. Inflation is often modeled by a scalar field 
with a self-interacting potential that gives rise to a slow-roll phase during 
which the energy density of the matter field remains nearly constant and 
the spacetime behaves like a quasi-de Sitter spacetime. However, there is no 
unique way of obtaining slow-roll inflation. There are numerous inflationary 
models that give rise to a quasi-de Sitter phase including 
single field inflation with various potentials, quasi-single field inflation, 
Dirac-Born-Infeld inflation and multi-field inflation. Fortunately, recent 
results from the Planck mission \cite{planck2013} and WMAP \cite{wmap7} are able to 
rule out various 
scenarios and favor a few. In particular, the data show that single field 
inflation with a quadratic potential is moderately disfavored and plateau-like 
inflationary potentials including the Starobinsky potential are favored.  
This has led to an increased interest in studying single field 
inflation with a Starobinsky potential. 

However, as shown by Borde, Guth and Vilenkin \cite{borde}, despite its great 
success the standard inflationary scenario, which is based on classical general 
relativity, is past incomplete due to the presence of big bang 
singularity. This is true for all models of scalar field inflation including 
the Starobinsky potential. The problem of a big bang singularity is an artifact 
of using Einstein's equations all the way to the Planck scale. It is expected 
that these problems will be resolved by a quantum theory of gravity which will 
depart from classical general relativity in the deep Planck regime. Several
important questions arise for such a theory: Is there a consistent extension of 
the inflationary scenario all the way to the Planck scale? Will inflation occur 
naturally, or would one require special fine tuning on the initial conditions to 
obtain a desired phase of inflation? To answer these questions the fundamental 
quantum gravity theory, while modifying the Planck scale physics, must be in 
agreement with classical general relativity (GR) when the spacetime curvature is 
well below the Planck scale. The theory will have to provide a consistent framework 
where one can study the evolution of scalar and tensor perturbations. 
Additionally, the theory should be able to provide natural initial conditions for 
the quantum perturbations in the deep Planck regime. Will these initial 
conditions lead to the usual Bunch-Davies state used in the standard inflationary 
scenario, or are they different? Can there be quantum gravity corrections to the 
standard power spectra? Modern CMB observations have put strong 
constraints on various aspects of the scalar and tensor power spectra. Are the 
predictions from the fundamental quantum gravity theory compatible with 
these observational data? As was shown in \cite{aan2,aan3} all these questions 
can be answered in the framework of loop quantum cosmology (LQC). In this paper 
we will study the inflationary scenario with a Starobinsky potential in the LQC 
framework.

Over the past decade, LQC has emerged as a concrete framework to address these issues 
and study the evolution of both the background and cosmological perturbations 
all the way to the Planck scale \cite{aps2,aps3,aan2,aan3} (for other approaches to studying cosmological 
perturbations within LQC, see \cite{FernandezMendez:2012vi,Gomar:2015oea,Barrau:2013ula,
Cailleteau:2011kr,Barrau:2014maa,Bojowald:2011aa, Zhu:2015ata}). Indeed, a key feature 
of LQC models is the resolution of classical big bang type singularities via a 
non-singular quantum bounce. Extensive analytical and numerical studies have been 
carried out to understand the nature of a variety of cosmological models in the 
quantum gravity regime where the spacetime curvature is Planckian. For example: 
the flat Friedmann-Lema\^{i}tre-Robertson-Walker (FLRW) model which we will be 
focusing on in this paper \cite{bojowaldprl01,aps2,aps3,acs,ps09}, open and 
closed FLRW models \cite{apsv,warsaw_closed,singhk-1}, the flat FLRW model in the 
presence of cosmological constants \cite{bp,kp1,ap}, various homogeneous and 
anisotropic models \cite{awe2,awe3,we1,b1madrid1,b1madrid2} and 
their generalizations to include inhomogeneities \cite{wemadrid1,bmp1,bmp2} 
(see \cite{as1} for an up-to-date review). Occurrence of the quantum bounce is 
a robust prediction of LQC in all these models which remains independent of the 
choice of initial conditions and energy conditions \cite{acs,ps09,dgs2,dgms,
cyclic}. 

It is now natural to raise questions discussed above in the context of LQC. 
Extensive analytical and numerical studies performed in \cite{aan2,aan3} show 
that the answer to most of the questions mentioned is in the affirmative for the
single field inflationary model with a {\it quadratic potential}. The Planck 
scale LQC correction greatly modifies the pre-inflationary dynamics while being 
in harmony with the inflationary phase \cite{asloan_prob2}.\footnote{The 
background dynamics in the pre-inflationary phase has also been 
investigated in great detail for power-law inflation \cite{rs} and non-minimally 
coupled scalar field \cite{Artymowski:2012is} in the flat FLRW model. 
For an inflationary scenario in the presence of anisotropies in LQC, 
see e.g. \cite{gs3}. 
} 
The altered behavior in the background due to UV quantum gravity corrections 
that resolves the singularity can leave imprints on the infrared modes of the 
cosmological perturbations \cite{aan3}. This seems counter-intuitive at first 
but, as shown in \cite{aan3} and also explained in \sref{sec:phenom}, is 
actually consistent. It is a remarkable feature of LQC that while alleviating 
the fundamental problem of a classical singularity one can obtain further 
interesting phenomenological consequences that can be compared against recent 
observations \cite{ag2,agulloassym,ashtekar_barrau}. 

%A detailed investigations of this interplay between the UV modes of the 
%background and the altered behavior of the infrared cosmological modes have 
%been performed for the simplest inflationary model: single field inflation with a 
%quadratic potential. 
%
How about inflation with a Starobinsky potential, 
which is in fact favored by the data?  A priori it is not obvious that any of 
the results that are true for the quadratic potential will also hold for the 
Starobinsky potential. For instance, the dynamics of both models is 
considerably different \textit{during} inflation resulting in the different 
predictions made by each model. 
So why would the pre-inflationary dynamics be similar? Furthermore, due to the 
existence of a maximum energy density in LQC, the initial data surface for the 
quadratic potential is a closed surface. In contrast, as a result of the flattening of 
the Starobinsky potential, in this model the initial data surface is open. Does 
this have any consequences? It was found that for the quadratic potential 
inflation was nearly inevitable: almost all the initial data lead to inflation 
compatible with observations \cite{asloan_prob2,ck1}.\footnote{For a 
detailed investigation of the initial conditions in the pre-bounce phase of 
this model, see \cite{Linsefors:2013cd}.} Since the initial data 
surfaces are completely different (even its topologies are distinct), one 
may wonder how likely the occurrence of inflation is for a Starobinsky 
potential and whether enough e-folds are generated for {\it any} of the 
initial conditions? In addition, what happens to the LQC corrections to scalar
and tensor power spectra? Are there any features that can distinguish between
the quadratic and Starobinsky potential? We summarized the main findings of our
analysis in a Letter \cite{bg1}. In this paper, we provide the details of the 
analysis and elaborate on the phenomenological investigation of the background 
spacetime and quantum perturbations.

In order to study the evolution of quantum perturbations on the LQC modified 
background geometry we used the framework of quantum field theory on quantum cosmological 
spacetime \cite{akl,aan2} and follow the strategy used in \cite{aan3,
agulloassym,ag2} to obtain potentially observable consequences of the quantum 
geometry for the Starobinsky potential. First, we focus on the dynamics of the 
background spacetime. Next, we study the quantum perturbations on this quantum 
modified background. Surprisingly, as shown in \cite{aan2}, the quantum 
perturbations on the quantum geometry experience a smooth, dressed geometry 
which encodes all the LQC quantum modifications relevant for the perturbations. 
Due the non-singular nature and finite maximum curvature at the bounce long 
wavelength modes of the perturbations are affected by the curvature. As a 
result their power spectra is different on large scales as compared to standard 
inflationary power spectra. For inflation with a quadratic potential, this can 
lead to some of the large scale anomalies observed in the CMB 
\cite{agulloassym,ag2}. 
 
The main results of our analysis are as follows. As expected, the classical 
big bang singularity is resolved via a quantum bounce and the energy density is 
maximum there.  This defines for us the space of initial conditions at the 
bounce. Along the lines of the analysis of the quadratic potential, using the 
recent observational data, we define a desired phase of slow roll for 
compatibility with observations. We find that, although the initial data 
surface is very different from that in the quadratic case, all initial 
conditions -except for a very tiny fraction- lead to the desired slow-roll 
phase in the future evolution. Interestingly, only the kinetic energy dominated bounces are compatible with observations and none of the potential energy  
dominated bounces undergo inflation. This is in contrast with the quadratic 
case, where all potential energy  dominated cases are compatible with 
observations.  Therefore, for observationally interesting initial conditions 
details of the Starobinsky potential do not matter and the background quantum 
gravity regime is practically the same as that for the quadratic case.\footnote{
Although for the quadratic case, all potential energy dominated initial 
conditions are compatible with observations, only the kinetic energy dominated cases are extensively studied in the literature (as these are more interesting 
from a LQC perspective as well as for computational limitations).  Therefore, 
when we refer to results for the quadratic potential, we in fact only mean to 
compare to the cases reported in the literature: the kinetic energy 
dominated bounces.} This results in similar prediction for the scalar and power 
spectra of the observable modes as for the quadratic case. Interestingly 
though, we find that there exist initial conditions for which the scalar power 
spectrum of super horizon modes do depend on the potential. The tensor 
spectrum, however, remains the same. This could lead to interesting 
implications for potential dependent correction to the slow-roll consistency 
relation and signatures of LQC on the CMB power spectrum due to non-Gaussian 
modulation of the power spectra as suggested in \cite{schmidthui,agulloassym}.

We organize the paper as follows. We begin by reviewing the key features of LQC 
in \sref{sec:lqc} and the framework of quantum fields on a quantum background in 
\sref{sec:qftqst}. In \sref{sec:initial}, we fix the mass parameter of the 
potential using recent observational results from the Planck mission and 
describe initial data for both the background and the perturbation. The  
results are described in \sref{sec:results}. We summarize our main 
results and discuss future extensions of this work in \sref{sec:disc}.

Throughout this paper, we use the following conventions. The spacetime metric 
has signature -+++. All numerical results will be quoted in Planck units ($G, \hbar$ 
and $c$ are all equal to one). 
However, we often write the Planck mass $\mpl:=\hbar^{1/2} c^{1/2} G^{-1/2}$ 
(or Planck length $\lp$ or Planck second $\spl:=\lp/c$)  explicitly to aid 
the reader and in equations we will display $G$ and $\hbar$ to highlight 
their physical content. The Planck energy density also plays an important 
role and is given by $\rhopl:=\hbar^{-1}c^7G^{-2}$. 
(Note that the Planck mass differs from the reduced Planck mass that is often used in cosmology by a factor of $\sqrt{8\pi}$.)

%%%%%%%%%%%%%%%%%%%%%%%%%%%%%%%%%%%%%%%%%%%%%%%%%%%%%%%%%%%%%%%%%%%%
\section{The framework}
In this section we first pay a brief visit to the main features of LQC 
relevant to our study. Then, we discuss the framework of 
quantum field theory on quantum spacetime that we will be
using to study the evolution of cosmological perturbations. We will be working
with the flat FLRW model with $\mathbb R^3$ spatial topology. 

%%%%%%%%%%%%%%%%%%%%%%%%%%%%%%%%%
\subsection{Features of LQC}
\label{sec:lqc}
LQC is based on the canonical quantization framework of loop quantum gravity (LQG).\footnote{
For other approaches to cosmology within the framework of canonical LQG see e.g. 
\cite{Achour:2014rja,Wilson-Ewing:2015lia,Bodendorfer:2015hwl}} 
The LQG quantization procedure is to first write the classical Hamiltonian 
constraint of GR in terms of connection and triad variables. Then, the holonomy of the
connection and the flux of the triads are promoted to quantum operators
(rather than the connection and the triads themselves). This quantization scheme 
is diffeomorphism invariant and predicts a discrete quantum geometry with a
minimum area gap given by $\Delta~\lp^2=4\sqrt{3}\pi\gamma\lp^2$ with 
$\gamma=0.2375$ the Barbero-Immirzi parameter whose value is fixed via 
black hole entropy calculation in LQG \cite{as1,Meissner_gamma,Domagala_gamma}. 
LQC applies the LQG techniques to cosmological models: one first identifies 
the underlying symmetries of the spacetime and writes the classical Hamiltonian
constraint in terms of the symmetry reduced connection and triad variables. The 
quantization procedure is then applied to these symmetry reduced variables. 
The resulting quantum Hamiltonian constraint in LQC is a quantum difference 
equation instead of a differential equation. The discreteness in this equation 
is fixed by the minimum area gap $\Delta$. 
The evolution of a wavefunction of the Universe,
governed by the quantum Hamiltonian constraint turns out to be
non-singular. That is, the physical wavefunction remains peaked on non-zero 
volumes and the expectation values of physical observables remain 
finite and non-singular throughout the evolution. Moreover, an initially sharply
peaked state, which describes a macroscopic universe at late times, remains
sharply peaked at all times; even in the deep Planck regime \cite{aps2,aps3,kp}.
Therefore, one can express the leading features of the loop quantum geometry by
tracking the ``mean geometry'' described by the expectation values of the
physical observables for a sharply peaked wavefunction. This leads to the so
called effective description of LQC.

The effective description of LQC is based on the geometric formulation of
quantum mechanics \cite{josh,schilling1}, where via a judicious choice of
semi-classical states and by looking at the expectation values of the physical
observables one can find a faithful embedding of the classical phase space into the
quantum phase space. The effective Hamiltonian obtained via this procedure gives
rise to modifications to the classical Friedmann equation \cite{vt}:
\be
\label{eq:fried}H^2 = \left(\f{\dot a}{a}\right)^2 = \f{8 \pi G}{3} \rho
\left(1-\f{\rho}{\rhomax}\right),
\ee
where $H$ is the Hubble parameter, $a$ the scale factor appearing in the FLRW models,
$\rho$ the energy density of the matter sources and $\rhomax$ is the maximum upper bound on the energy density:
\be
 \rhomax = \f{18 \pi}{\Delta^3~G~\lp^2}\approx 0.41~\rhopl.
\ee
It is important to note that $\rhomax$ is a universal constant whose origin is
solely quantum geometric in nature, as evident by the factor $\Delta$ in the
equation above. 
Similarly a modified Raychaudhuri equation can also be derived:
\be
\label{eq:ray}\f{\ddot a}{a}  =  -\f{4 \pi G}{3} \rho
\left(1-\f{4\rho}{\rhomax}\right) -{4 \pi G} P \left(1-\f{2\rho}{\rhomax}\right),
\ee
where as usual $P$ refers to the pressure of the matter field(s).
This can also be written as:
\be
\label{eq:hubray} \dot H = -4\pi G \left(\rho + P\right)
                            \left(1-\f{2\rho}{\rhomax}\right).
\ee 
In obtaining the modified Friedmann and Raychaudhuri equation 
no assumptions about the matter field has been made. Thus, the non-singular
features of LQC remain to hold for all types of matter field. Furthermore,
the form of the conservation equation remains the same as in the 
classical theory:
\be
 \dot{\rho} + 3 H \left(\rho + P\right) = 0.
\ee
If the matter source is a scalar field $\phi$ with a standard kinetic term and 
a potential $V(\phi)$, the conservation equation above is equivalent to the 
evolution equation of $\phi$:
\be
	\ddot{\phi} + 3 H \dot{\phi} + \f{\partial V(\phi)}{\partial \phi} = 0
\ee
where we used that $\rho=\dot{\phi}^2/2 + V(\phi)$ and $P=\dot{\phi}^2/2 - V(\phi)$. 
Note that the quantum geometry only influences the dynamics of the scalar 
field through the Hubble rate $H$. 

Before moving on to the next subsection let us briefly remark on the main aspects
of the modified dynamics:
\begin{itemize}
\item In contrast to the classical Friedmann equation where $H^2$ is directly proportional to the energy density, the modified Friedmann equation 
(\eref{eq:fried}) contains a correction term quadratic in $\rho$ that appears with 
an important negative sign. This implies that for a matter field which
satisfies the standard energy conditions, in LQC the Hubble rate (and 
hence $\dot a$) vanishes when $\rho=\rhomax$ even though $H$ in this scenario 
can never vanish in GR! Moreover, \eref{eq:ray} dictates that the second time 
derivative of $a$ is positive there, consequently $a$ attains a minimum. This 
is the point where the scale factor undergoes a bounce. 

\item In the limit when the spacetime curvature in very small compared to the
Planckian value, the quadratic corrections on the right hand side of \eref{eq:fried} and
\eref{eq:ray} are negligible and one recovers the classical Friedmann and
Raychaudhuri equations. Therefore, during the backward evolution of a large
classical universe, LQC and classical evolution trajectories coincide as long as
$\rho \ll \rhomax$. In the quantum regime, when $\rho$ becomes a few percent of
$\rhomax$, the two theories start to deviate. If one continues
the evolution further back in time, all classical trajectories undergo a 
big-bang singularity and all curvature scalars diverge, while the LQC 
trajectories bounce at a finite scale factor. 

\item It is straightforward to conclude from \eref{eq:fried} and (\ref{eq:ray})
that the Hubble rate also has an upper maximum when $\rho=\rhomax/2$:
\be
  H_{\rm max} = \sqrt{\f{1}{4 \gamma^2 \Delta}} = 0.93\,\mpl. 
\ee
Unlike in the classical theory, where $H$ monotonically approaches infinity 
during the approach to the big-bang singularity, in LQC $H$ increases to 
$H_{\rm max}$ and then quickly falls to zero at the bounce. For kinetic energy
dominated bounces the duration between the bounce and $H=H_{\rm max}$ is of the
order of $0.2 \spl$. 
 
\item It has been explicitly shown for a flat FLRW model sourced with a 
perfect fluid with constant equation of state that the upper bounds on 
the energy density and Hubble rate necessarily lead to an upper bound on all other
curvature scalars \cite{ps09}. Hence, all strong curvature singularities are generically
resolved. Similar robustness results have also been obtained in the presence of
anisotropies in the Bianchi-I spacetime \cite{ps11}.
\end{itemize}

%%%%%%%%%%%%%%%%%%%%%%%%%%%%%%%%%
\subsection{Quantum field theory on quantum spacetime}
\label{sec:qftqst}
In this subsection, we will briefly discuss the framework of quantum field
theory on a quantum spacetime background. For details and an application to
an inflationary scenario with a quadratic potential see \cite{akl,aan2,aan3}.
In the standard treatment of cosmological perturbations, one treats 
cosmological perturbations as quantum fields on a classical FLRW background 
geometry that solves Einstein's equations. 
In a quantum gravity theory however, since the background
is no longer given by a classical metric, one needs a framework of quantum fields
propagating on {\it quantum spacetime}. Such a framework was developed in the
context of LQC in \cite{akl}. Compared to the framework of quantum field theory 
in curved background, quantum fields in quantum spacetime seems an extremely 
difficult problem at first: Now one has to keep track of the evolution of the 
quantum geometry (described by a wavefunction $\Psi (a,\phi)$) rather than just 
a few time dependent parameters that describe the metric. To be strict, 
the situation is even worse: 
in the quantum theory there is no notion of a metric. 
Nevertheless, as described in great detail in
\cite{akl,aan2}, surprising simplifications occur if the cosmological
perturbations can be treated as test fields. That is, if quantum fields
describing the perturbations evolve without affecting the evolution of the
background wavefunction or, in other words, the backreaction of these 
perturbations is negligible.

In this test field approximation, the dynamics of the quantum field describing
cosmological perturbations on the background quantum geometry is equivalent to 
propagation of quantum fields on a quantum modified effective geometry described 
by a dressed metric $\tilde{g}_{ab}$ \cite{aan2,aan3}:
\be
\label{eq:dressed}
  \tilde{g}_{ab}~dx^a~dx^b = \tilde{a}^2 \left(-d\tilde\eta^2 +
d\vec{x}^2\right),
\ee
where $\tilde a$, the dressed scale factor, and $\tilde\eta$, the dressed conformal
time, are:
\be
\label{eq:atilde}
  \tilde a = \left( \f{\langle \widehat{H}_0^{-1/2} \, \widehat{a}^4 \, \widehat{H}_0^{-1/2}\rangle}
{\langle \widehat{H}_0^{-1} \rangle} \right)^{1/4} \qquad {\rm and} \qquad 
   d\tilde\eta = \langle \widehat{H}_0^{-1/2} \rangle \left(\langle \widehat{H}_0^{-1/2} \, \widehat{a}^4 \, \widehat{H}_0^{-1/2}\rangle\right)^{1/2} d\phi,
\ee
where $\widehat{H}_0$ is the background Hamiltonian and the expectation values are taken 
with respect to the background quantum geometry state given by $\Psi (a, \phi)$. 

It is important to note that the dressed metric does not describe the 
full quantum geometry, but rather the quantum modified smooth geometry that 
is relevant for cosmological perturbations. Moreover, $\tilde a$ is not merely
the expectation value of the scale factor operator. As apparent from the
expressions in the above equation, the dressed quantities know about the
background quantum geometry and hence about the quantum fluctuations in 
the background geometry. In the derivation of the dressed metric approach no
assumptions were made regarding the type of the background state $\Psi$, 
therefore this framework is valid for an arbitrary physical state that solves 
the quantum Hamiltonian constraint of LQC. In this paper, we will consider 
very sharply peaked states for which the relative volume dispersion 
$\Delta V/V \ll 1$.  This leads to a further computational simplification, 
namely, the dressed scale factor can now be well approximated by the scale 
factor of the effective dynamical equations (\eref{eq:fried}) discussed in the 
previous subsection.

The cosmological perturbations on this background are described by the gauge-invariant Mukhanov-Sasaki scalar mode $\hat{\mathcal{Q}}$ and two tensor modes $\hat{\mathcal{T}}^{(1)}$ and $\hat{\mathcal{T}}^{(2)}$. It is simplest to work in the (comoving) momentum space, where the mode functions satisfy a simple equation mode by mode. We decompose $\hat{\mathcal{Q}}$ as
\be 
 	\hat{\mathcal{Q}}(\tilde{\eta},\vec{x}) = \int \! \f{\d^3 k}{(2\pi)^3} \left( \hat{a}_{\vec{k}} \, q_k (\tilde{\eta}) + \hat{a}_{-\vec{k}}^{\dagger} \, q_k^* (\tilde{\eta}) \right) e^{i \vec{k} \cdot \vec{x}},
\ee 
where $k=|\vec{k}|$ and the mode functions $q_k (\tilde{\eta})$ are assumed to be square-integrable. The mode functions are solutions to
\be 
	\label{eq:scalar}
 	q''_k (\tilde{\eta}) + 2 \f{\tilde{a}'}{\tilde{a}} q'_k(\tilde{\eta}) + \left(k^2 + \tilde{\mathcal{U}} \right) q_k(\tilde{\eta})=0
\ee
where primes denote derivatives with respect to the dressed conformal time $\tilde{\eta}$ and with 
\be 
	\tilde{\mathcal{U}} = \f{\langle H_0^{-1/2} \, \widehat{a}^2 \, \widehat{\mathcal{U}}(\phi) \, \widehat{a}^2 \, H_0^{-1/2}\rangle}
{\langle H_0^{-1/2} \, \widehat{a}^4 \, H_0^{-1/2} \rangle}  \qquad {\rm and} \qquad
	\mathcal{U}(\phi) = a^2 \left( \mathfrak{f} \, V(\phi) - 2 \sqrt{\mathfrak{f}} \,
	 \f{\partial V}{\partial \phi} + \f{\partial^2 V}{\partial \phi^2} \right),
\ee
where $\mathfrak{f}= 12 \pi G \, \f{\f{1}{2}\dot{\phi}^2}{\f{1}{2}\dot{\phi}^2 + V(\phi)}$.
Note that the effective potential $\tilde{\mathcal{U}}$ is completely determined by the background quantities. Furthermore, the mode functions are normalized so that their Klein-Gordon norm is $i$ and, consequently, the creation and annihilation operators satisfy the standard commutation relations, $[\hat{a}_{\vec{k}}, \hat{a}^\dagger_{\vec{k}'} ] = \hbar \, (2\pi)^3 \,   \delta^{(3)}(\vec{k}-\vec{k}')$. A similar decomposition can be done for the tensor perturbations. The main difference is that the tensor mode functions $e_k(\tilde{\eta})$ now satisfy an even simpler equation than the scalar modes:
\be 
	\label{eq:tensor}
	e''_k (\tilde{\eta}) + 2 \f{\tilde{a}'}{\tilde{a}} e'_k(\tilde{\eta}) + k^2 			e_k(\tilde{\eta})=0.
\ee
These equations for the scalar and tensor modes, \eqref{eq:scalar} and
\eqref{eq:tensor}, will play an essential role in the discussion of power
spectra and quantum gravity effects on very long wavelength modes.

%%%%%%%%%%%%%%%%%%%%%%%%%%%%%%%%%%%%%%%%%%%%%%%%%%%%%%%%%%%%%%%%%%
\section{Parameters and initial conditions}
\label{sec:initial}
As discussed in the previous sections, here we are interested in studying the
LQC extension of the inflationary scenario where inflation is driven by a scalar
field $\phi$ with a self interacting potential given by \cite{Barrow:1988xi,Starobinsky:2001xq,DeFelice:2010aj}:
\be
	\label{eq:pot}
	V(\phi) = \f{3M^2}{32 \pi G} \left(1- e^{-\sqrt{\f{16\pi G}{3}} \phi}
\right)^2,
\ee
where $M$ is the mass of the scalar field. The above potential is also known in
the literature as the Starobinsky potential.
In this section, we use the most recent observational data from the Planck 
mission \cite{planck15xx} to fix this mass parameter $M$. We then determine the 
values of slow-roll parameters that are necessary for compatibility with 
observations. This analysis is relevant for the next section where we determine 
the set of initial conditions for the background quantities that is compatible 
with observations. Finally, we end this section with a discussion of the 
initial conditions for the quantum perturbations.

\subsection{Background initial data}
\label{sec:backinitial}
Let us first recall the definition of the slow-roll parameters. There are 
various definitions that are useful in different settings. Here, we will use 
the following two distinct sets of slow-roll parameters:
\begin{itemize}
  \item the \textit{(Hubble) slow-roll parameters}, which are defined in terms of $H$ and its derivatives. This is an infinite tower of parameters, however, here we will only need the first two, which are given by
       \be
         \epsilon = -\f{\dot H}{H^2}\, \qquad {\rm and} \, 
                     \qquad \delta = -\f{\ddot H}{2 \dot H H}.
       \ee      
  \item and the\textit{ potential slow-roll parameters}, which are defined for single field inflationary models in terms of the potential of the inflaton field and its derivatives. The first two are
       \be 
       \epsilonV = \frac{1}{16 \pi G} \left(\frac{V'}{V}\right)^2  
                         \qquad {\rm and} \, \qquad 
       \deltaV = \frac{1}{8 \pi G} \frac{V''}{V} .            
        \ee
\end{itemize}
Within the slow-roll approximation, i.e. neglecting terms that are quadratic in the slow-roll parameters, these parameters are related in the following way
	\begin{align*}
	\epsilon &\simeq \epsilonV, \\
	\delta &\simeq \deltaV - \epsilonV,
	\end{align*}
where $\simeq$ is to indicate that the equality is true only within the slow-roll approximation.

The spectral index $\ns$ and the field amplitude $\as$ of the scalar perturbations at the time when the mode $\ks$ exits the Hubble radius during inflation are \cite{planck15xx}
\begin{align*}
&\ns = 0.9645\pm0.0062, \\
&\as = (2.474 \pm 0.116) \times 10^{-9}.
\end{align*}
Note that here we are using $\ks=0.002$ Mpc$^{-1}$ while Planck reports $\ns$ 
and $\as$ at $k=0.05$ Mpc$^{-1}$. This choice for $\ks$ was made for the 
following two reasons: 
(i) the value of $\Ns$ used in this paper is computed for $0.002$ Mpc$^{-1}$
\cite{planck15xx}, and
(ii) for an easy comparison with the previous investigations of the inflationary 
scenario in LQC \cite{aan3}. We use the power law scalar powerspectrum 
with constant $\ns$ to compute $\as$ and $\ns$ at $\ks=0.002$ Mpc$^{-1}$ from
the Planck data.

For a quadratic potential, $\ns$ and $\as$ uniquely specify  
$\epsilon$, $\delta$, $H$, $\phi$ and $\dot{\phi}$ at the time $\ts$ when $\ks$ 
exited the horizon. Additionally, it determines the inflaton mass $m$ 
within the slow-roll 
approximation, which is consistent with observational error bars. However, for 
the Starobinsky potential $\delta \neq 0$, unlike for the quadratic potential. 
Therefore, $\ns$ and $\as$ are not enough and one needs one more piece of 
information to uniquely specify these parameters. We take as additional input 
$\Ns$, which is the number of e-folds from the time when $\ks$ exited the 
horizon to the end of slow-roll inflation defined by $\epsilon=1$. We will 
refer to this period as the \textit{desired slow roll phase of inflation}. For 
Starobinsky inflation, $\Ns\in (54,62)$. Here, we will use the middle value 
$\Ns = 58$.\footnote{The number of e-folds during the desired slow-roll phase 
as determined by the Planck data depends on the particular model of inflation. 
Thus, different inflationary models predict different $\Ns$ given the same 
Planck data, for instance,  $\Ns \simeq 56$ for a quadratic potential.} 

Now having $\ns$, $\as$ and $\Ns$ at our disposal we use Einstein's equations,
the relations among $\as$, $\ns$, the Hubble rate and the slow roll parameters to 
solve for $\epsilon, \delta, H, \phi$ and $\dot{\phi}$ at the time of horizon 
crossing as well as to fix the mass parameter $M$ appearing in the Starobinsky 
potential. This gives us the following complete system of six equations and 
six unknowns:
\begin{align}
&\as  = \f{H_*^2}{\pi \epsilon_* \mpl^2} \label{eq:as} \\
%&\ns -1 = -2 \epsilon_1 - \epsilon_2 \simeq -4\epsilon + 2\delta  \label{eq:ns} \\
&\ns -1  \simeq -4\epsilon_* + 2\delta_*  \label{eq:ns} \\
&\Ns \simeq -1.04 + \frac{3}{4} e^{\sqrt{\frac{16 \pi G}{3}} \phi_*} - \sqrt{3 \pi G} \phi_*  \label{eq:nstar}\\
& 3 H \dot{\phi_*} + V'(\phi_*) \simeq 0 \label{eq:eomphi} \\
& H^2 = \frac{8\pi G}{3} \left(\frac{1}{2} \dot{\phi_*}^2 + V(\phi_*) \right) \label{eq:friedmann}\\
& \dot{\phi_*}^2 (\epsilon_* - 3) + 2 \epsilon_* V(\phi_*) = 0 \label{eq:fromdefeps}
\end{align}
where a prime denotes the derivative with respect to $\phi$ and it is understood
that all conditions are to be evaluated at the time $\ts$. The first equation
relates the amplitude of the scalar power spectrum $\as$ to the Hubble rate $H$ 
and slow-roll parameter $\epsilon$ at horizon crossing and the second one is a 
relation true for a standard single field inflationary model with arbitrary 
potential. To obtain the third equation, we used
$\Ns := \ln \frac{a_{\rm end}}{a(\ts)} \simeq - 8 \pi G 
        \int_{\phi_*}^{\phi_{\rm end}} \frac{V}{V'}\d \phi$ 
and approximated that $\epsilon \simeq \epsilonV$ at the end of 
inflation from which we obtain that $\phi_{\rm end}=0.187~\mpl$. The fourth and 
fifth equations are the two independent Einstein's equations for a homogeneous 
and isotropic universe, where for the first one $\ddot{\phi}$ is neglected 
which follows from the slow-roll approximations. The last equation is a 
rewriting of the definition for $\epsilon$ after substitution of Einstein's 
equations. This set of equations has a unique solution for $M>0$, yielding the 
following values when $\ks$ crosses the horizon
\begin{align}
\epsilon_* &= 1.98 \times 10^{-4} \qquad \, \, \, \, \phi_* = 1.080 \notag \, \mpl \\
\delta_* &= - 1.73 \times 10^{-2} \qquad \dot{\phi}_* = - 4.80 \times 10^{-9} \, \mpl^2 \label{eq:values} \\
H_* &= 1.21 \times 10^{-6} \, \mpl \notag
\end{align}
and the mass parameter in the potential is
\be
	M=2.51 \times 10^{-6} \, \mpl.
\ee
Note that the desired phase of slow-roll requires all of the values in 
\eref{eq:values} to be obtained at the time of horizon crossing. The Starobinsky 
potential with the mass here obtained is shown in \fref{fig:pot}. It is 
bounded below by zero: $V(\phi) \geq 0$. This has an important consequence, 
because the total matter density is bounded by $\rhomax$, 
$|\dot{\phi}|$ is now bounded above by $\sqrt{2~\rhomax}$. 
On the positive side, i.e. $\phi\rightarrow -\infty$, the potential tends 
to a finite value $V\rightarrow \f{3 M^2}{32 \pi} \approx 10^{-12}$, whereas 
for $\phi \to - \infty$ the potential diverges.

Just as for FLRW models in GR, the space of initial data for the effective 
description of LQC is four dimensional. It consists of the values of the scale 
factor $a$, the Hubble rate $H$, $\phi$ and $p_{\phi}=a^3 \dot{\phi}$ given at
the initial time, which then give a unique solution to the effective equations. 
The value of the scale factor at the initial time has a constant rescaling freedom
which leaves the physical results unaltered. Utilizing this freedom we fix the
scale factor at the bounce $a_{\rm B}=1$ without loss of any generality.
We give our initial data at the bounce which is characterized by the following
two properties: 
(i) the Hubble rate at the bounce $H_{\rm B}=0$ as the scale factor has a minimum there, 
and 
(ii) the energy density $\rho=\rhomax=:\phidb^2/2+V(\phib)$ from \eref{eq:fried}. 
The first condition fixes the Hubble rate at the bounce and the second 
condition implies that specification of $\phib$ at the bounce determines 
$\phidb$ upto a sign. 
Hence, with $a_{\rm B}$ fixed, the value of $\phib$ and the sign of $\phidb$ 
completely determine the initial data and thus are the only free 
parameters. (Note that this is slightly different from the case with the 
quadratic potential, where $\phib$ can be considered to be the only free 
parameter, since solutions with a different sign of $\phidb$ can easily be 
obtained by using the symmetry of the potential.)

The maximum of the energy density at the bounce restricts the space of initial 
data from the entire real line to $[\phi_{\rm min}= - 3.47 \, \mpl ,\infty)$. Note 
that this is different from inflation with a quadratic potential $V(\phi) = 
m^2 \phi^2/2$ where the initial data surface is compact ($|\phib|< \sqrt{2 \rhomax/m}$) 
because the potential grows unbounded for both negative and positive $\phi$. A 
meaningful way to divide this space of initial data is by the kinetic and 
potential energy in the scalar field at the bounce. This is quantified by the 
polytropic index of the scalar field at the bounce $w_B=(\phidb^2 - V(\phib))/
(\phidb^2 + V(\phib))$: the bounce is considered kinetic energy dominated if 
$|w_{\rm B} -1| <10^{-3}$ and potential energy  dominated if $|w_{\rm B} + 1| 
< 10^{-3}$. We will refer to the initial conditions with $w_{\rm B}=1$ 
($w_{\rm B}=-1$) as {\it extreme} kinetic (potential) energy dominated.

\bfig
\includegraphics[width=0.6\textwidth]{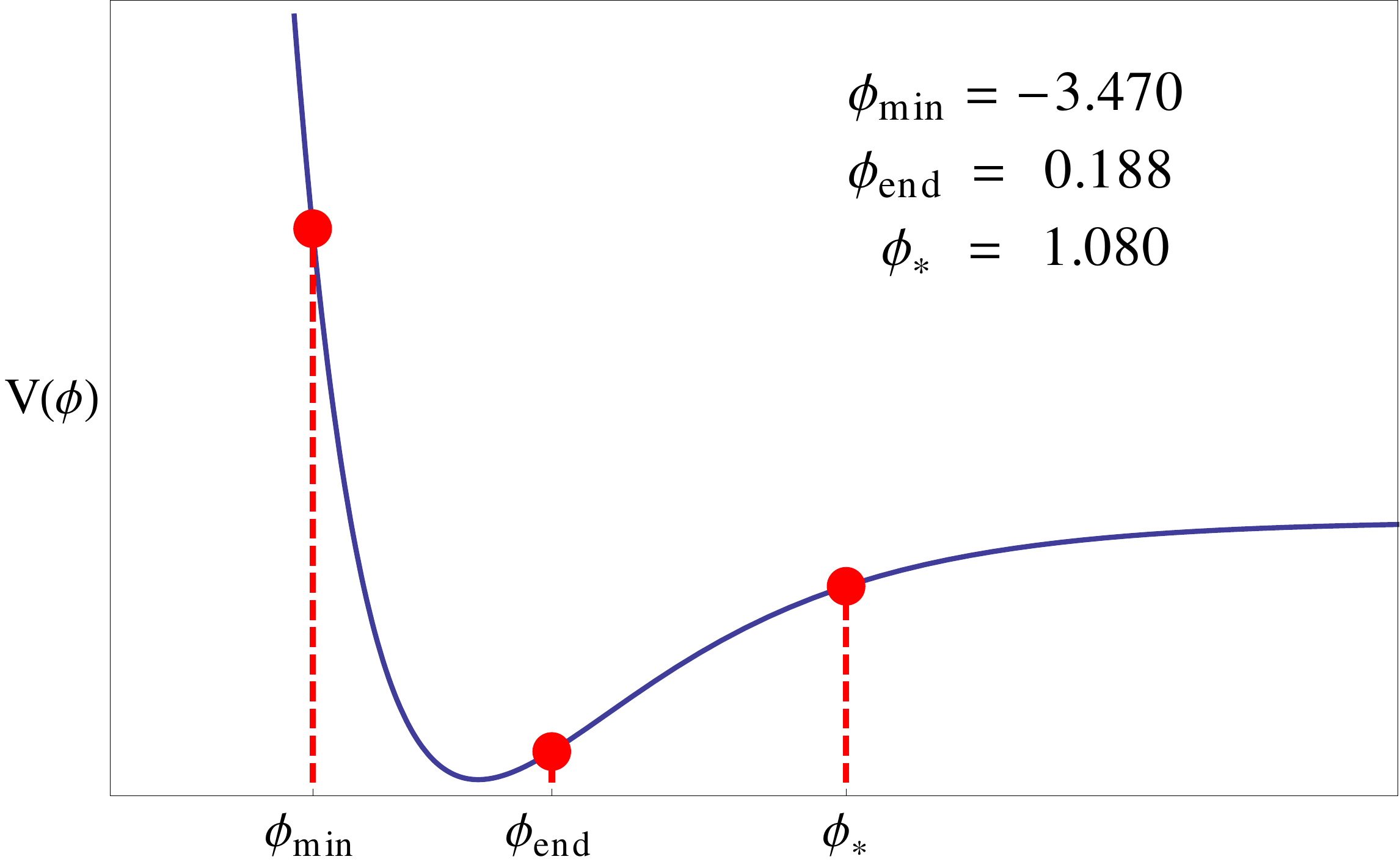}
\caption{Plot of the Starobinsky potential. The potential is bounded below by
zero and for positive $\phi$ the potential is also bounded above, $V\rightarrow
\f{3 M^2}{32 \pi G} \approx 10^{-12}~\rhopl$, while for negative $\phi$ the 
potential is unbounded from above. The energy density at the bounce restricts 
$\phib$ to $[\phi_{\rm min},\infty)$. The desired slow-roll phase starts when
$\phi=\phi_*$ and $\dot{\phi}<0$ and ends when $\phi=\phi_{\rm end}$.}
\label{fig:pot}
\efig

\vskip0.5cm
\noindent {\bf Remark:} In order to compute the mass parameter $M$, we have 
used observational data and assumed the standard power-law power spectrum 
obtained from standard inflation where perturbations are taken to be in 
Bunch-Davies state at the onset of inflation. As we will see later in this 
paper, however, in LQC modes are not in a Bunch-Davies state at the onset of 
inflation and the power spectrum is not exactly the same as the standard one. 
There are corrections for small $k$ modes while there is agreement for large 
$k$. Due to these corrections, our computation of $M$ is internally inconsistent. 
Strictly speaking, one should use the true LQC power spectrum in order to compute the 
mass parameter. This requires significant numerical work to perform analysis 
over the whole parameter space, which has been recently done for the quadratic 
potential in \cite{agullomorris}. They find that, while using the LQC power 
spectrum to compute $M$ is conceptually important, it leads to very small 
corrections in the power spectra and the main results remain unchanged. We expect 
a similar conclusion for the Starobinsky potential. Nevertheless, it is an important 
issue and we leave it for future investigations. In this paper, we bypass this 
issue by choosing initial conditions for which the reference mode corresponds 
to large enough $k$ where the Bunch-Davies and LQC power spectra differ 
very little. Therefore, any correction to $M$, if any, would be too small to 
impact our results.

\subsection{Initial states for quantum perturbations}
In de Sitter spacetime one can construct a unique vacuum state for linear, 
cosmological perturbations that are regular and respect the symmetries of the 
de Sitter background. These vacuum states are the so-called Bunch-Davies 
states. In the standard inflationary scenario the spacetime metric can be 
approximated by a de Sitter metric at the onset of slow-roll. Therefore, it is 
a reasonable assumption that the observable modes are in the Bunch-Davies state 
there. In LQC, on the other hand, the modified pre-inflationary dynamics 
extends all the way to the Planck scale where the background is very different
from de Sitter spacetime. Therefore, quantum perturbations cannot be chosen to 
be in a Bunch-Davies state close to the bounce. As suggested in \cite{aan2,aan3,ana}, one 
can still use the physical principles behind the construction of Bunch-Davies states to 
obtain vacuum states in LQC. These principles require that the states should be 
ultraviolet {\it regular} and {\it invariant under the symmetries} of the 
underlying spacetime, which in our case is given by the dressed metric 
$\tilde g_{ab}$ (\sref{sec:qftqst}). This leads us to a 4th order adiabatic 
state at bounce. It turns out that if the states are chosen to be 4th order
adiabatic initially, they remain so throughout the evolution. Another physical 
input used in the choice of states stems from the {\it test field 
approximation}. That is, the backreaction of the quantum perturbations on the 
background spacetime in quantum gravity regime is negligible. This is essential 
for the self-consistency of the dressed metric approach. 

Unlike in de Sitter spacetime, however, this procedure does not single out a 
unique vacuum state. In this paper, we will use four different types of states, 
which are based on different physical principles:
\begin{itemize}
\item {\it Type I}: this vacuum state is constructed such that it leads to
suppression of power at large scales in the CMB as suggested in \cite{ag2};
\item {\it Type II}: this vacuum state minimizes the stress energy tensor of the
perturbations to zero at some initial time (called `instantaneous vacuum' in \cite{ana}); 
\item {\it Type III}:  this vacuum state is the approximate WKB solution of the evolution equations at some initial time (called `obvious vacuum' in \cite{aan3}); and
\item {\it Type IV}: this vacuum state is a combination of the type II vacuum for large $k$ and a Minkowski-like vacuum for small $k$ as suggested in \cite{agulloassym}. 
\end{itemize} 
Note that these initial states are not specified at the bounce, but slightly before the bounce (specifically, at $t= -10 \, \spl$) for technical reasons that arise for infrared modes. 

%%%%%%%%%%%%%%%%%%%%%%%%%%%%%%%%%%%%%%%%%%%%%%%%%%%%%%%%%%%%%%%%%%
\section{Results}
\label{sec:results}
Let us now turn to the numerical results. Recall that according to the framework
of quantum field theory on quantum spacetime, as long as the perturbations can be treated as test fields, the
information about the background geometry that is relevant for the evolution of
perturbations can be encoded into a dressed effective geometry. The dressed
metric is obtained from the expectation values of {\it only} certain moments 
of the scale factor and field momentum with respect to the background 
wavefunction $\Psi(a,\phi)$ \eref{eq:atilde}. One does not need all the 
details of the quantum fluctuations of the wavefunction in order to determine 
the behavior of the quantum perturbations.\footnote{It is worth emphasizing that
the dressed metric is just a mathematical tool to study the evolution of
perturbations. It does not exactly describe the background quantum geometry 
which is still given by the background wavefunction that contains all the 
information of the true background geometry.} Strictly speaking, given a 
state $\Psi$ in the physical Hilbert space, the dressed (denoted with $tilde$ 
on top) and the effective scale factor are different from each other. However, 
as shown via explicit numerical simulations in \cite{ag1}, the numerical 
differences between the dressed metric and the effective metric is very small 
(less than a percent) for sharply peaked states ($\Delta V/V \ll 1$). In this 
paper, we consider such sharply peaked states and approximate the dressed 
geometry with the effective geometry discussed in \sref{sec:lqc}.

The remainder of this section is divided into three subsections. In the first, 
we explore the evolution of the background effective geometry for various 
classes of initial conditions. We find the subset of initial conditions that 
leads to the desired slow-roll phase, which is necessary for compatibility with 
observations. As expected, the occurrence of the quantum bounce is a generic 
feature of the loop quantum geometry and the desired slow-roll phase happens
for almost all of the initial conditions. In the second subsection, we describe
why the pre-inflationary dynamics matters for perturbations and how the
ultraviolet quantum gravity effects can lead to particle production for infrared
modes of perturbations. Finally in the third subsection, we present 
the numerical evolution of the scalar and tensor perturbations on this 
effective background geometry, compute their power spectra at the end of 
inflation, and extract the window of initial conditions relevant for 
observational consequences of pre-inflationary dynamics of LQC. Evolution of 
the background geometry in the quantum gravity regime, and that of the quantum 
perturbations, turns out to be mostly similar to that for the quadratic 
potential studied in \cite{aan3}. However, there is a small subset of initial
conditions for which Starobinsky potential leaves different signatures for very
long wavelength modes.

\bfig
\ig[width=0.6\textwidth]{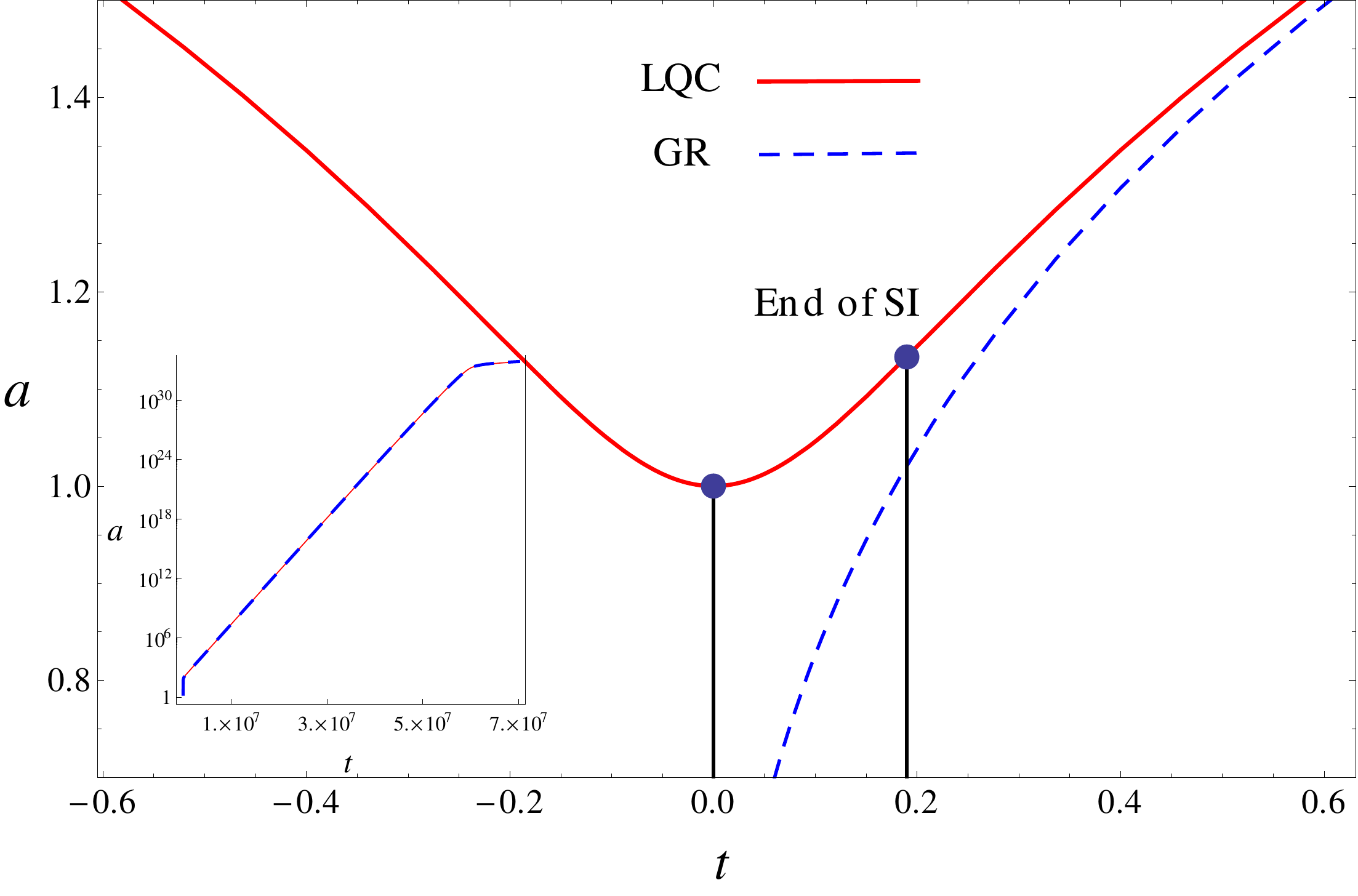}
\caption{Evolution of the scale factor with $\phib=-1.45$ and $\phidb>0$. The
(red) solid curve and the (blue) dashed curve respectively show the evolution
of the scale factors in LQC and classical GR. The bounce happens at $t=0$ where the
classical trajectory goes to big bang singularity. Following the bounce, there
is short phase of super-inflation which ends at $t\sim0.19\,\tpl$. In the future
evolution LQC and classical GR trajectory converge. The inset shows the late
time evolution of the LQC scale factor undergoing inflation which ends at
$t\sim6\times10^7\,\tpl$.}
\label{fig:aback}
\efig
\bfig[tbh]
\ig[width=0.49\textwidth]{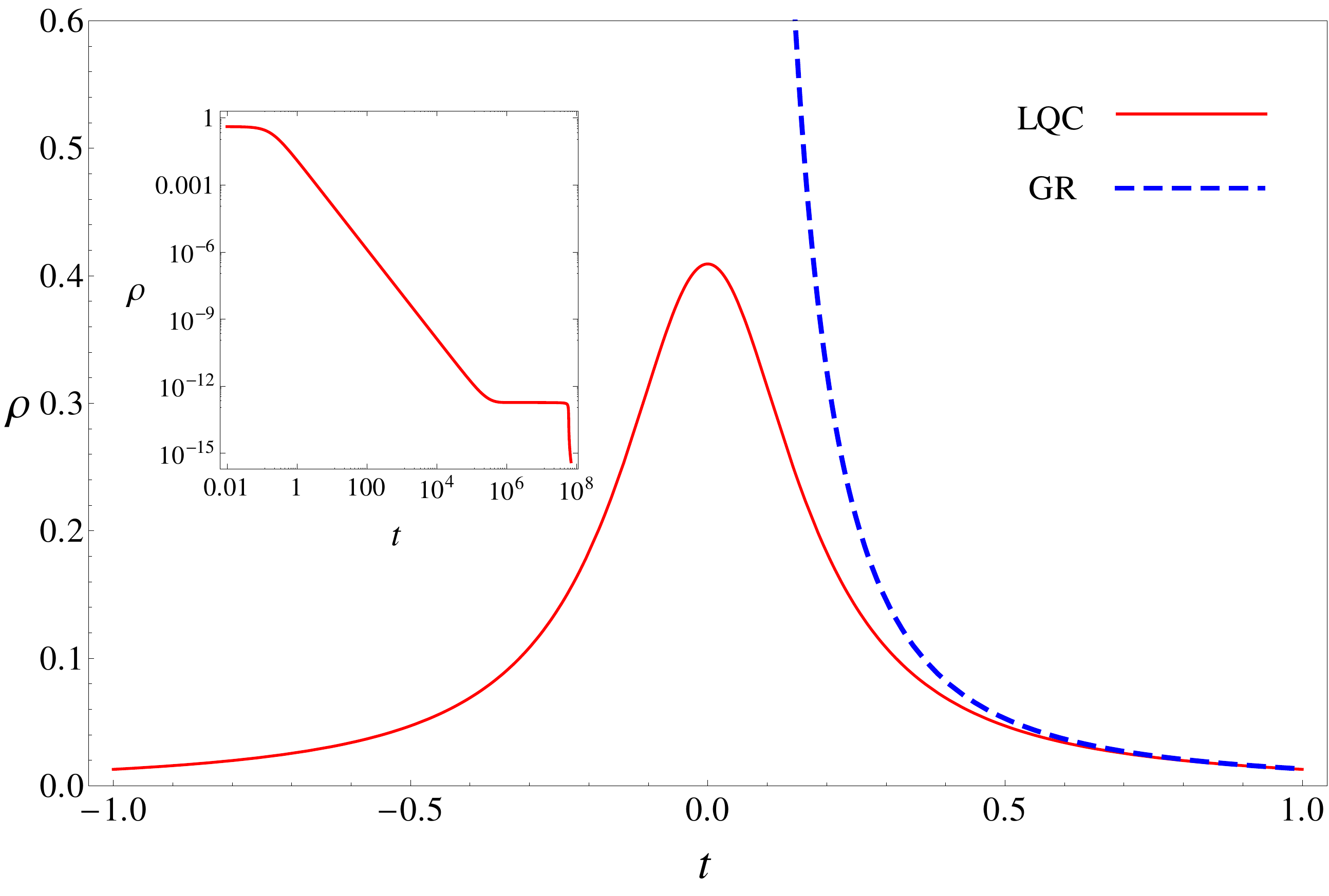}
\hskip0.2cm
\ig[width=0.49\textwidth]{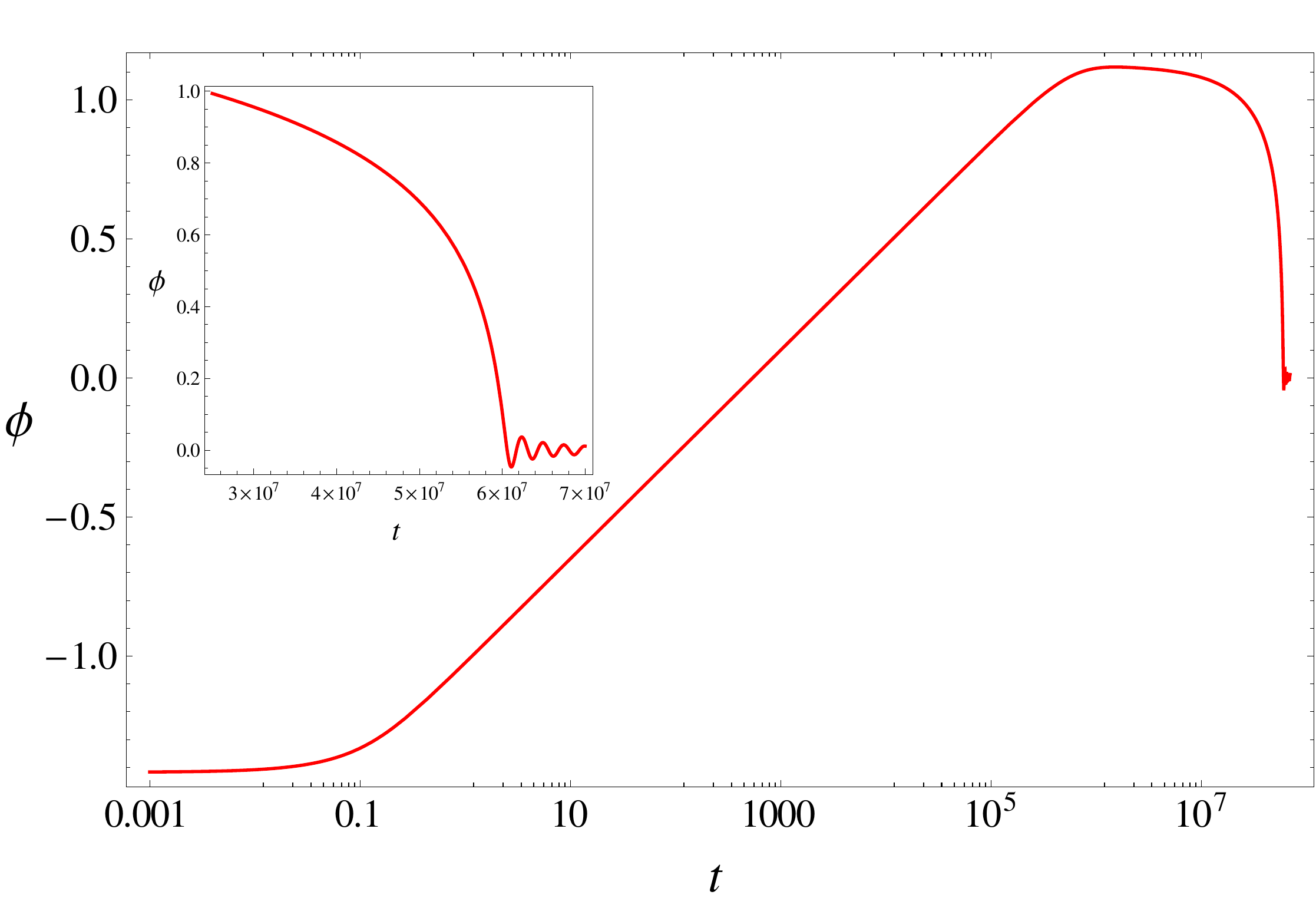}
\caption{The energy density of the matter field (left panel) and the evolution
of the scalar field (right panel) for the same initial conditions as in
\fref{fig:aback}. The energy density in LQC has a maximum at the bounce
while in classical GR it diverges. The scalar field evolves from
negative values to the positive valued slow-roll side of the potential. The
insets show the late time evolution of $\rho$ and $\phi$. During inflation the
energy density is of the order of $10^{-12}\,\rhopl$ and remains almost
constant. At the end of inflation there is a drop in $\rho$ while the scalar
field oscillates at the minimum of the potential around $\phi=0$.} 
%It is
%apparent from the figures that the LQC and classical GR deviate only in the
%quantum gravity regime close to the bounce, and quickly start to co-incide with
%each other away from the bounce.}
\label{fig:rhoback}
\efig

\subsection{Background evolution}
\label{sec:resultback}
\fref{fig:aback} shows the evolution of the scale factor in LQC (solid curve) 
and classical GR (dashed curve) for a representative case with 
$\phib=-1.45~\mpl$ and $\phidb>0$. The corresponding energy density and the scalar 
fields are shown in \fref{fig:rhoback}. From these plots, it is immediately 
obvious that in the vicinity of the quantum bounce, which is well inside the 
Planckian regime, there are significant deviations between classical GR and LQC:
First, the scale factor bounces from a non-zero value in LQC while in the 
classical theory $a\rightarrow0$.\footnote{As described in \sref{sec:backinitial}, 
we have chosen the parameters so that the scale factor at the bounce is 
$a_{\rm B}=1$. This choice was made for the convenience of the calculations, 
but carries no physically relevant information. Physical observables for both 
the background and perturbations are independent of this choice.}
Second, the energy density is finite at the bounce for the LQC evolution, but 
diverges in classical GR. Following the quantum bounce, there is a brief phase 
($\sim 0.2\,\tpl$) of faster than exponential accelerated expansion, called 
super-inflation. During this phase the Hubble rate quickly grows from zero to 
its maximum value. At the end of super-inflation, characterized by $H=H_{\rm
max}$ (and, consequently, $\dot{H}=0$), the energy density of the scalar field 
is half of that at the bounce (see \eref{eq:hubray}). This super-inflationary 
phase is the most quantum gravity dominated part of the evolution. In further 
evolution, the Hubble rate, the energy density and hence the spacetime curvature
decrease monotonically while the scale factor grows. After approximately a 
hundred Planck seconds, the spacetime curvature falls well below the Planckian 
value where LQC and classical GR are in excellent agreement with each other. 
The desired phase of slow roll inflation begins typically when the energy 
density becomes of the order of $10^{-12}\,\rhopl$. Recall that by slow roll, 
we mean the period between when $\ks$ exited the horizon to when the first 
Hubble slow-roll parameter becomes equal to one. At the onset of this slow-roll 
phase $\phi$ and $\dot{\phi}$ attain the values in \eqref{eq:values}. Note that 
by definition the number of e-folds during this slow-roll phase of inflation is 
58 for the Starobinsky potential, but the total number of e-folds during the entire 
inflationary epoch (when $\ddot{a}>0$) can be much higher.

The right panel of \fref{fig:rhoback} shows the corresponding evolution of the  
scalar field with the same initial conditions. The scalar field starts from the
left side of the minimum of the potential (see \fref{fig:pot}) with a positive 
$\phidb$ due to which it rolls down the potential. As the further evolution 
takes place, the field crosses the minimum and starts climbing up on the 
slow-roll side and at some point in the evolution crosses $\phi_*$ for the first
time. But this time the field has a positive velocity and does not have the
correct slow-roll parameters. Due to the remaining kinetic energy, the field
climbs a little higher up the potential until KE=0, and that is when the field
turns around and starts rolling down the potential with $\dot\phi<0$. In what
follows, the field crosses $\phi_*$ once more, this time with $\dot\phi<0$ and
the slow-roll conditions are met. When $\phi$ becomes 0.19 $\mpl$ the slow-roll 
phase ends and the field continues to roll down the potential and finally 
oscillates at the minimum of the potential around $\phi=0$ as shown in the 
inset of the right plot in \fref{fig:rhoback}. Note that, if the field did not 
have enough kinetic energy to climb high enough the potential and cross 
$\phi_*$ during the climb, the slow-roll conditions would not have been 
obtained.

Let us now study the evolution of the background geometry in some more detail 
for a variety of cases by giving the initial conditions at the bounce. 
We will divide the set of initial conditions into two classes: 
(i) positive inflaton velocity ($\phidb>0$) and 
(ii)  negative inflaton velocity ($\phidb<0$), 
and track the evolution of $\phi$, $\dot\phi$ and $a$ for different $\phib$.
As discussed in \sref{sec:initial}, the range of $\phib$ is semi-infinite: 
$\phib\in(\phi_{\rm min},~\infty)$ where $\phi_{\rm min}=-3.47~\mpl$ 
corresponds to the extreme potential energy dominated bounce with zero kinetic 
energy and $\phib\rightarrow\infty$ to the extreme kinetic energy dominated 
bounce with zero potential energy. As we will see in this section -- 
irrespective of the sign of $\phidb$ -- the potential energy dominated bounces 
never give rise to the desired phase of slow-roll. This is in contrast to the 
quadratic potential where the potential energy  dominated cases are in fact the 
ones with large $\phib$ resulting into a huge amount of slow-roll inflation.

\subsubsection{Positive inflaton velocity: $\phidb>0$}
Let us begin by discussing the evolution of the Ricci curvature scalar $\mathcal{R} $ 
in the quantum gravity regime. The behavior of the Ricci scalar is critical in understanding the evolution of the quantum perturbations, as the Ricci scalar determines the length scale at which the properties of the curved background become important and the quantum perturbation can no longer be treated as fields propagating on a flat background. The expression for the Ricci scalar in terms of the
Hubble rate $H$ is:
\be
 \mathcal R = 6\left(\dot H + 2 H^2\right),
\ee
where the dot represents the derivative
with respect to the proper time. Now substituting the expressions of $H$ and its
derivative in terms of the energy density of the matter field, we have:
\be
  \mathcal R = -24 \pi G \, \rho(1 + w) \left(1- 2 \f{\rho}{\rhomax}\right) + 24
\pi G \, \rho \left(1- \f{\rho}{\rhomax}\right),
\ee
where, as before, $w$ is the polytropic index appearing in the equation of state of the matter field. It is clear from the
above equation that specifying the energy density of the matter field and its equation
of state completely determines the Ricci curvature. At the bounce, where
$\rho=\rhomax$, we have:
\be
  \mathcal R_{\rm B} = 24 \pi G \, \rhomax \, (1+w_{\rm B}) 
                     = \f{432~\pi^2}{\Delta^3~\lp^2} (1+w_{\rm B}). 
\ee
In this way, while the maximum of the energy density at the bounce is fixed at
$\rhomax$, the value of the Ricci scalar at the bounce depends on the type of matter field
under consideration. For example, for dust $\mathcal R_{\rm
B}^{\rm (dust)} = \f{432~\pi^2}{\Delta^3~\lp^2}$, for radiation $\mathcal R_{\rm
B}^{\rm (rad)} = \f{576~\pi^2}{\Delta^3~\lp^2}$ and for a stiff fluid 
--which is equivalent to a massless scalar field-- 
$\mathcal R_{\rm B}^{\rm (stiff)} = \f{864~\pi^2}{\Delta^3~\lp^2}$.

\bfig
\ig[width=0.50\textwidth]{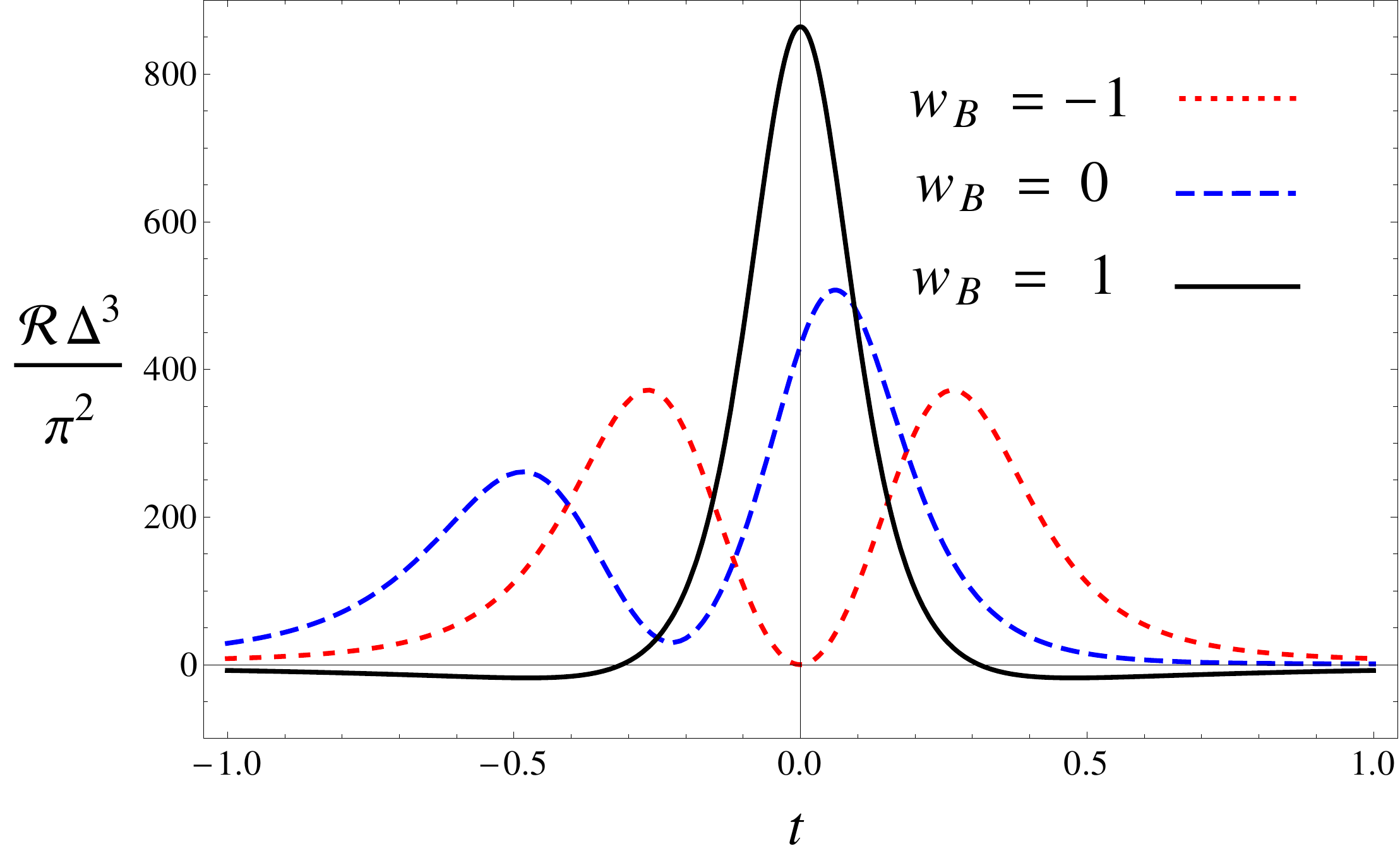} 
\hskip0.2cm
\ig[width=0.47\textwidth]{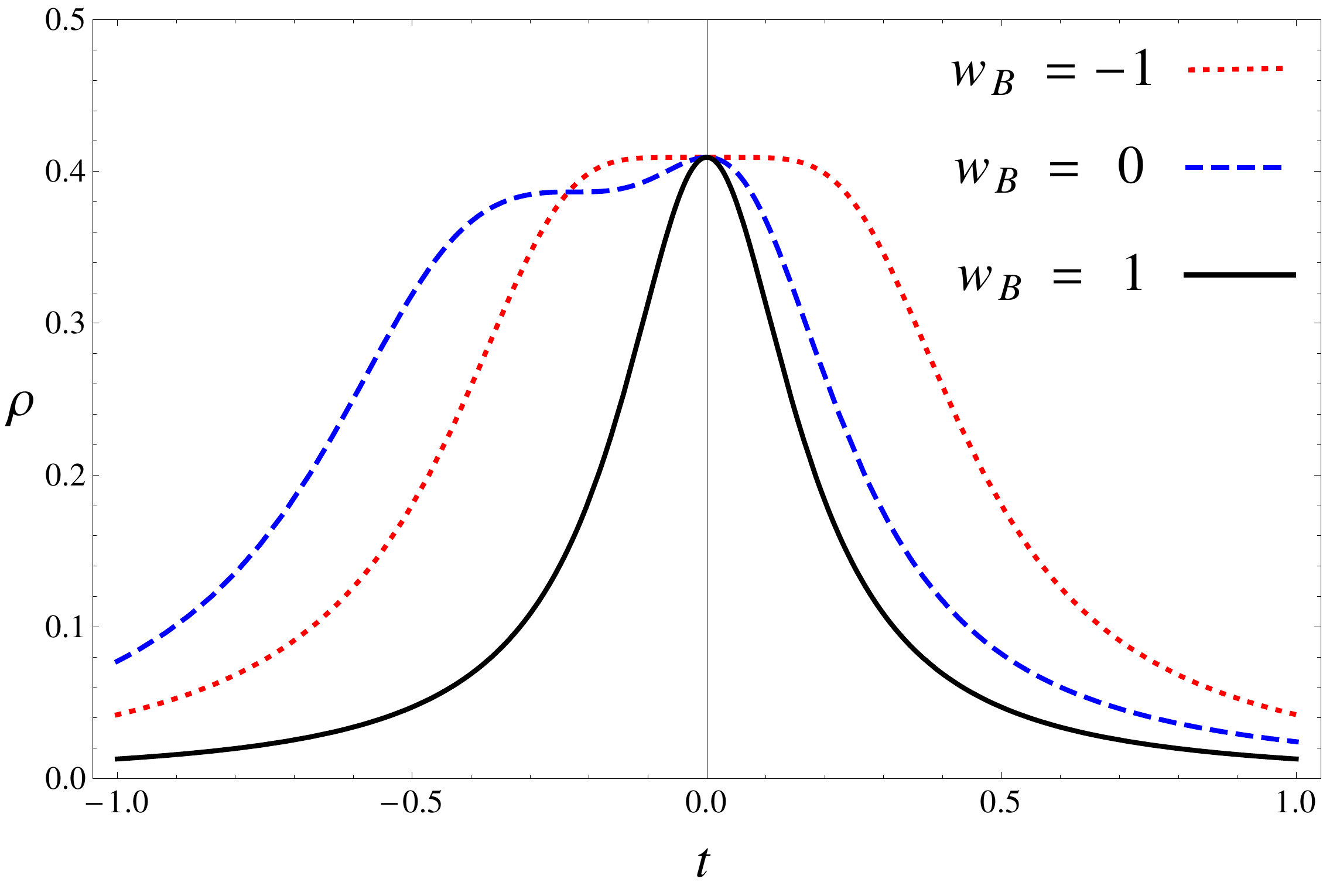} 
\caption{Plots of the Ricci scalar and energy density near the bounce (left and right panel, respectively). For the extreme potential energy  
dominated case ($w_{\rm B} = -1$), the Ricci scalar vanishes at the bounce, whereas for the extreme kinetic energy dominated case ($w_{\rm B} = 1$), the Ricci scalar is maximum at the bounce. The energy density is maximum at the bounce for all cases. 
For the intermediate case when the potential and kinetic energy are exactly equal at the bounce ($w_{\rm B} = 0$), 
the Ricci scalar $\mathcal{R}_{\rm B}$ is half of the maximum value attained for $w_{\rm B}=1$. For all conditions
$w_{\rm B}<1$, the Ricci scalar behaves non-monotonically in the future evolution. 
Moreover, the bounce is time-reversal symmetric for $w_{\rm B}=1$ and $-1$, while asymmetric for all other initial conditions. It is also apparent that the 
energy density falls more quickly for kinetic energy dominated cases than potential energy  
dominated ones. Hence, the quantum gravity regime last longer for potential energy  dominated bounces.}
\label{fig:ricci}
\efig

{\tiny
\begin{table}
\caption{
	Dynamical evolution for various $\phib$ with $\phidb>0$.	
	The value of $\phi$, $w$, $t$, $H$, $\epsilon$ and the total number 
of e-folds since the bounce $N$ is shown for various events. In particular, the events considered are the bounce, end of super-inflation (SI), moment when the kinetic energy equals the potential energy (KE=PE), onset of the desired slow-roll phase (onset) and  end of slow roll inflation (end). The zeros in the table are in fact $\mathcal{O}(\epsilon_{machine})$, but from analytic considerations we know these values are identically zero. The last three rows all satisfy $w_{\rm B}-1 <10^{-7}$.
}
\label{tab:posphid}
\begin{ruledtabular}
%ccccccc
\begin{tabular}{lllllll}
Event & $\phi$  & $w$ & t & $H$ & $\epsilon$  & N \\
\hline
Bounce & -3.47 &   -1  &  0   & 0  & $\infty$ & 0 \\
End SI & -3.25 &  0.32 & 0.46 & 0.93 & 0 & 0.20  \\
KE=PE  & -0.55 &   0   & $4.7 \times 10^{4}$ &$1.5 \times 10^{-5}$ & 1.5 & 8.12\\
Onset  &   NA  &   NA  & NA   & NA & NA & NA \\
End    &   NA  &   NA  & NA   & NA & NA & NA \\

\hline
Bounce & -3.39 & 0 & 0 & 0 & $\infty$ & 0 \\
End SI & -3.21 & 0.52 & 0.26 & 0.93 & 0 & 0.16 \\
KE=PE  & -0.54 & 0 & $4.7 \times 10^{4}$ & $1.5 \times 10^{-5}$ & 1.5 & 7.87 \\
Onset  & NA & NA  & NA & NA & NA & NA \\
End    & NA & NA  & NA & NA & NA & NA \\

\hline
Bounce & -1.49 & 1.0 & 0 & 0 & $\infty$ & 0 \\
End SI & -1.35 & 1.0 & 0.18 & 0.93 & 0 & 0.12 \\
KE=PE  & 0.91 & 0 & $2.4 \times 10^{5}$ & $1.7 \times 10^{-6}$ & 1.5 & 4.74 \\
Onset  & NA & NA  & NA & NA & NA & NA \\
End    & NA & NA  & NA & NA & NA & NA \\

\hline
Bounce & -1.45 & 1.0 & 0 & 0 & $\infty$ & 0 \\
End SI & -1.31 & 1.0  & 0.18 & 0.93 & 0 & 0.12  \\
KE=PE  & 0.95 & 0 & $2.4 \times 10^5$ & $1.7 \times 10^{-6}$ & 1.5 & 4.74 \\
Onset  & 1.08 & -1.0 & $2.7 \times 10^6 $ & $1.2 \times 10^{-6}$ & $1.9 \times 10^{-4}$ & 7.90\\
End    & 0.19 & -0.67 & $5.2 \times 10^7$ & $7.4 \times 10^{-7}$ & 0.50 & 67.14 \\

\hline
Bounce & -1.41 & 1.0 & 0 & 0 & $\infty$ & 0 \\
End SI & -1.27 & 1.0  & 0.18 & 0.93 & 0 & 0.12  \\
KE=PE  & 0.99 & 0 & $2.4 \times 10^5$ & $1.7 \times 10^{-6}$ & 1.5 & 4.74 \\
Onset  & 1.08 & -1.0 & $1.2 \times 10^7 $ & $1.2 \times 10^{-6}$ & $1.9 \times 10^{-4}$ & 19.35\\
End    & 0.19 & -0.67 & $6.1 \times 10^7$ & $7.4 \times 10^{-7}$ & 0.50 & 78.59 \\

%\hline
%Bounce %	& 0 & 1.0 & 1.0 & 0 & $\infty$ & 0 \\ %End SI
%	& 0.14 & 1.0  & 0.18 & 0.93 & 0 & 0.12  \\
%KE=PE 
%	& 2.40 & 0 & $2.3 \times 10^5$ & $1.7 \times 10^{-6}$ & 1.5 & 4.74 \\
%Onset
%	& 1.08 & -1.0 & $1.9 \times 10^10 $ & $1.2 \times 10^{-6}$ & $1.9 \times 10^{-4}$ & $2.44 \times 10^4$\\
%\red{End}
%	& 0.19 & ? & ? & ? & ? & ?
\end{tabular}
\end{ruledtabular}
\end{table}
\par}

As a result, this is where one of the main distinctions between the potential and kinetic
energy dominated bounces appears. For  kinetic energy dominated bounces, $w_{\rm B} \approx 1$ and
consequently $\mathcal R_{\rm B}^{\rm (kin)} \approx
\f{864~\pi^2}{\Delta^3~\lp^2}$. On the other hand, for potential energy  dominated bounces, when the kinetic energy is 
close to zero, $w_{\rm B} \approx -1$ and $\mathcal R_{\rm B}^{\rm (pot)} \approx 0$. 
That is, the Ricci curvature at the bounce is maximized for kinetic energy dominated bounces, but vanishes for potential energy  dominated bounces. 
Thus, for positive potentials, while the energy density always 
saturates its maximum at the bounce, the Ricci scalar does not. 
As a result, the evolution of the Ricci scalar to the future of the bounce is
not necessarily monotonic, unlike the evolution of the energy density. In 
\fref{fig:ricci}, we show the evolution of the Ricci scalar for kinetic and 
potential energy  dominated as well as for an intermediate case with $w_{\rm B}=0$ at the 
bounce. It is clear that for $w_{\rm B}=-1$ (extreme potential energy  domination) the Ricci 
curvature vanishes at the bounce, then attains a local maximum and falls 
in the future evolution. For $w_{\rm B}=1$ (extreme kinetic energy domination) the Ricci 
curvature is maximum at the bounce and fall monotonically in the future 
evolution. It is also interesting to see that the maximum value that the Ricci scalar attains in its future evolution for 
$w_{\rm B}=-1$ is less than that for $w_{\rm B}=1$. Hence, the curvature scalars for 
different initial conditions are drastically different close to the bounce in 
the quantum gravity regime. This is conceptually important as \textit{this implies that 
the modes of quantum perturbations will be excited in qualitatively different 
ways depending on whether the bounce is potential or kinetic energy dominated.} Furthermore, the energy density falls of slower for potential energy  dominated cases and therefore the quantum gravity regime lasts longer. However, it turns out that this interesting conceptual difference is not relevant for 
observations since the potential energy  dominated bounces do not lead to enough e-folds 
during slow-roll and thus are not compatible with observations. 

For our numerical simulations, we used the Runge-Kutta numerical integration 
scheme in {\it Mathematica 10} and performed over a hundred simulations. 
\tref{tab:posphid} shows some of the representative simulations for various
$\phib$ when the initial velocity of the inflaton is positive. We are interested
in the following events during the evolution: (i) bounce, (ii) end of 
super-inflation, (iii) equivalence of kinetic and potential energy 
(specifically, $w=0$), (iii) onset of the desired slow-roll phase and 
(iv) end of inflation. For each initial condition, we compute the following 
values at these events: $\phi$, $w$, $t$, $H$, $\epsilon$ and the total number 
of e-folds since the bounce $N$.  The first two initial conditions in the table 
correspond to an extreme potential energy  dominated case ($w_{\rm B}=-1$) and 
an equal distribution between kinetic and potential energy ($w_{\rm B}=0$). The 
remaining three entries correspond to initial conditions for which the kinetic 
energy dominates over the potential energy by a factor of $10^7$ or more.
Duration of super-inflation is largest for the first row, slightly smaller for
the second row. For the rest of the entries in the table the evolution remains
almost the same till KE=PE. This happens because in the first two rows, 
potential does affect the evolution near the bounce, while for the rest of the 
cases shown in the table the field behaves almost like a massless scalar and 
potential becomes relevant after KE=PE. It is apparent from the table that the 
total number of e-folds for $\phib\in(\phi_{\rm min}, -1.45)$, which are 
potential energy  dominated initial conditions, is less than 60. Hence, the 
desired slow-roll marked by $\epsilon_*=0.00019$ and $\delta_*=-0.017$ is not 
obtained. On the other hand, for $\phib \geq -1.45$ (which is kinetic energy 
dominated) the desired phase of slow-roll is necessarily contained in the 
evolution.  Clearly, the potential energy  dominated bounces are not compatible 
with observations. The reason for this is that these potential energy dominated 
bounces do not have enough initial kinetic energy to climb up the potential on 
the other side and reach $\phi_*$.

{\tiny
\begin{table}
\caption{
	Dynamical evolution for various $\phib$ with $\phidb<0$. The table is structured 
	the same way as \tref{tab:posphid}: Shown are the values of $\phi$, $w$, $t$, $H$, 
	$\epsilon$ and the total number of e-folds since the bounce $N$ for several events.
	In particular, the events considered are the bounce, end of super-inflation (SI),
	when the kinetic energy equals the potential energy (KE=PE), at the onset of the
	desired slow-roll phase (onset) and at the end of slow roll inflation (end). The
	first row corresponds to an initial condition with equal amounts of potential and
	kinetic energy, the second row has $w_{\rm B}=1$ and the remaining four rows satisfy
	$w_{\rm B}-1<10^{-12}$.
}
\label{tab:negphid}
\begin{ruledtabular}
%ccccccc
\begin{tabular}{lllllll}
Event & $\phi$  & $w$ & t & $H$ & $\epsilon$  & N \\

\hline
Bounce & -3.39 & 0 & 0 & 0  & $\infty$ & 0 \\
End SI & -3.46 & -1.0 & 0.22 & 0.93 & 0 & 0.07  \\
KE=PE  & -0.55 & 0 & $4.7 \times 10^{5}$ &$1.5 \times 10^{-5}$ & 1.5 & 8.46\\
Onset  & NA & NA  & NA & NA & NA & NA \\
End    & NA & NA  & NA & NA & NA & NA \\

\hline
Bounce & 0 & 1 & 0 & 0  & $\infty$ & 0 \\
End SI & -0.14 & 1.0 & 0.18 & 0.93 & 0 & 0.12  \\
KE=PE  & -1.43 & 0 & $6.2 \times 10^{2}$ &$6.2 \times 10^{-4}$ & 1.5 & 2.73\\
Onset  & NA & NA  & NA & NA & NA & NA \\
End    & NA & NA  & NA & NA & NA & NA \\

\hline
Bounce & 3.00 & 1.0 & 0 & 0 & $\infty$ & 0 \\
End SI & 2.86 & 1.0 & 0.18 & 0.93 & 0 & 0.12 \\
KE=PE  & 0.59 & 0 & $2.6 \times 10^{5}$ & $1.6 \times 10^{-6}$ & 1.5 & 4.77 \\
Onset  & NA & NA  & NA & NA & NA & NA \\
End    & NA & NA  & NA & NA & NA & NA \\

\hline
Bounce & 3.63 & 1.0 & 0 & 0 & $\infty$ & 0 \\
End SI & 3.49 & 1.0  & 0.18 & 0.93 & 0 & 0.11  \\
KE=PE  & 1.23 & 0 & $2.4 \times 10^5$ & $1.7 \times 10^{-6}$ & 1.5 & 4.74 \\
Onset  & 1.08 & -1.0 & $2.7 \times 10^6 $ & $1.2 \times 10^{-6}$ & $2.0 \times 10^{-4}$ & 7.84\\
End    & 0.19 & -0.69 & $5.2 \times 10^7$ & $7.4 \times 10^{-7}$ & 0.50 & 67.07 \\

\hline
Bounce & 3.67 & 1.0 & 0 & 0 & $\infty$ & 0 \\
End SI & 3.52 & 1.0  & 0.18 & 0.93 & 0 & 0.12  \\
KE=PE  & 1.27 & 0 & $2.4 \times 10^5$ & $1.7 \times 10^{-6}$ & 1.5 & 4.74 \\
Onset  & 1.08 & -1.0 & $1.2 \times 10^7 $ & $1.2 \times 10^{-6}$ & $1.9 \times 10^{-4}$ & 19.37\\
End    & 0.19 & -0.69 & $6.1 \times 10^7$ & $7.4 \times 10^{-7}$ & 0.50 & 78.61 \\
\hline
Bounce & 4.00 & 1.0 & 0 & 0 & $\infty$ & 0 \\
End SI & 3.86 & 1.0  & 0.18 & 0.93 & 0 & 0.12  \\
KE=PE  & 1.60 & 0 & $2.3 \times 10^5$ & $1.8 \times 10^{-6}$ & 1.5 & 4.74 \\
Onset  & 1.08 & -1.0 & $1.9 \times 10^8 $ & $1.2 \times 10^{-6}$ & $1.9 \times 10^{-4}$ & 239\\
End    & 0.19 & -0.69 & $2.4 \times 10^8$ & $7.4 \times 10^{-7}$ & 0.50 & 298 \\
\end{tabular}
\end{ruledtabular}
\end{table}
\par}

\bfig
\ig[width=0.50\textwidth]{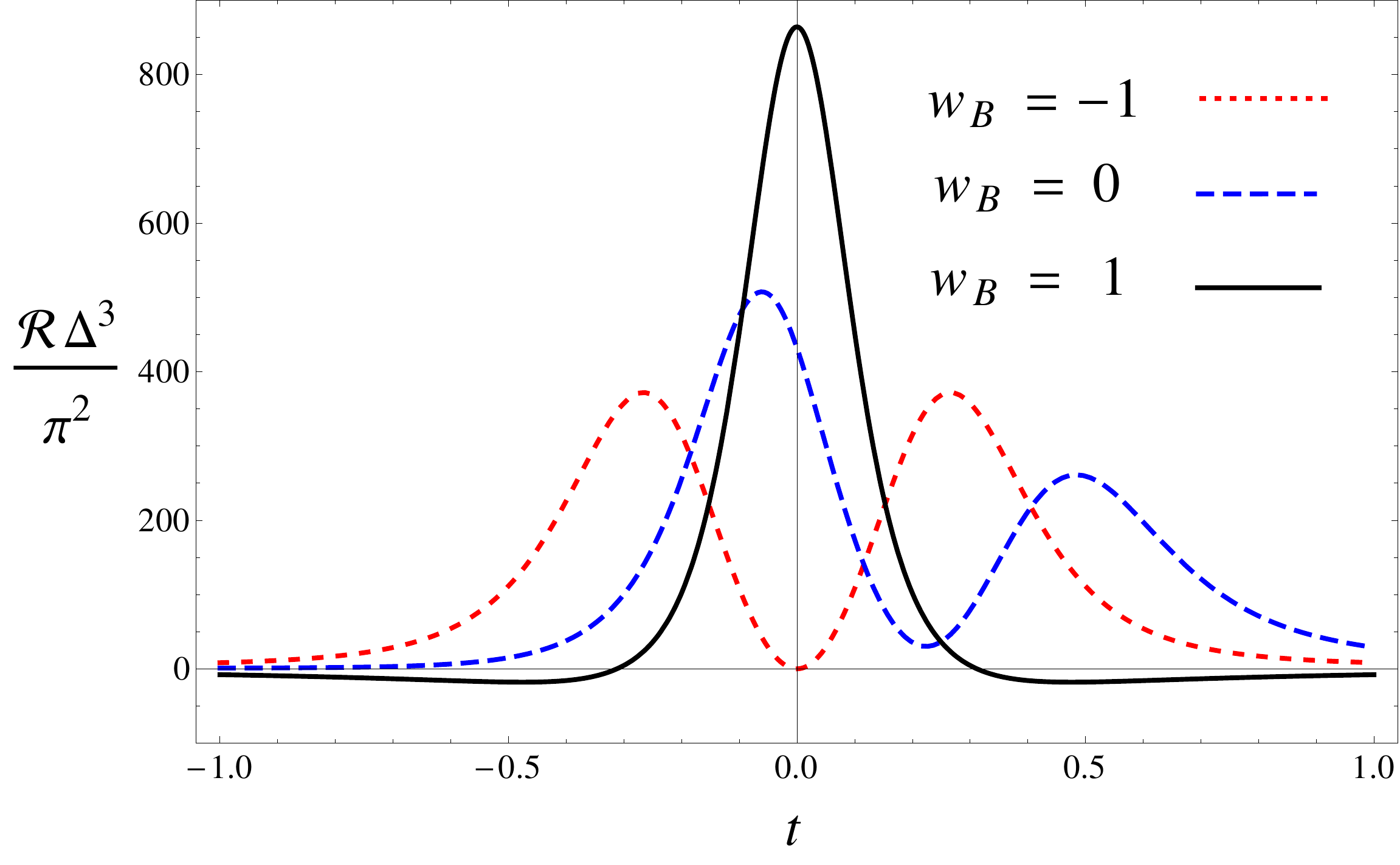} 
\hskip0.2cm
\ig[width=0.47\textwidth]{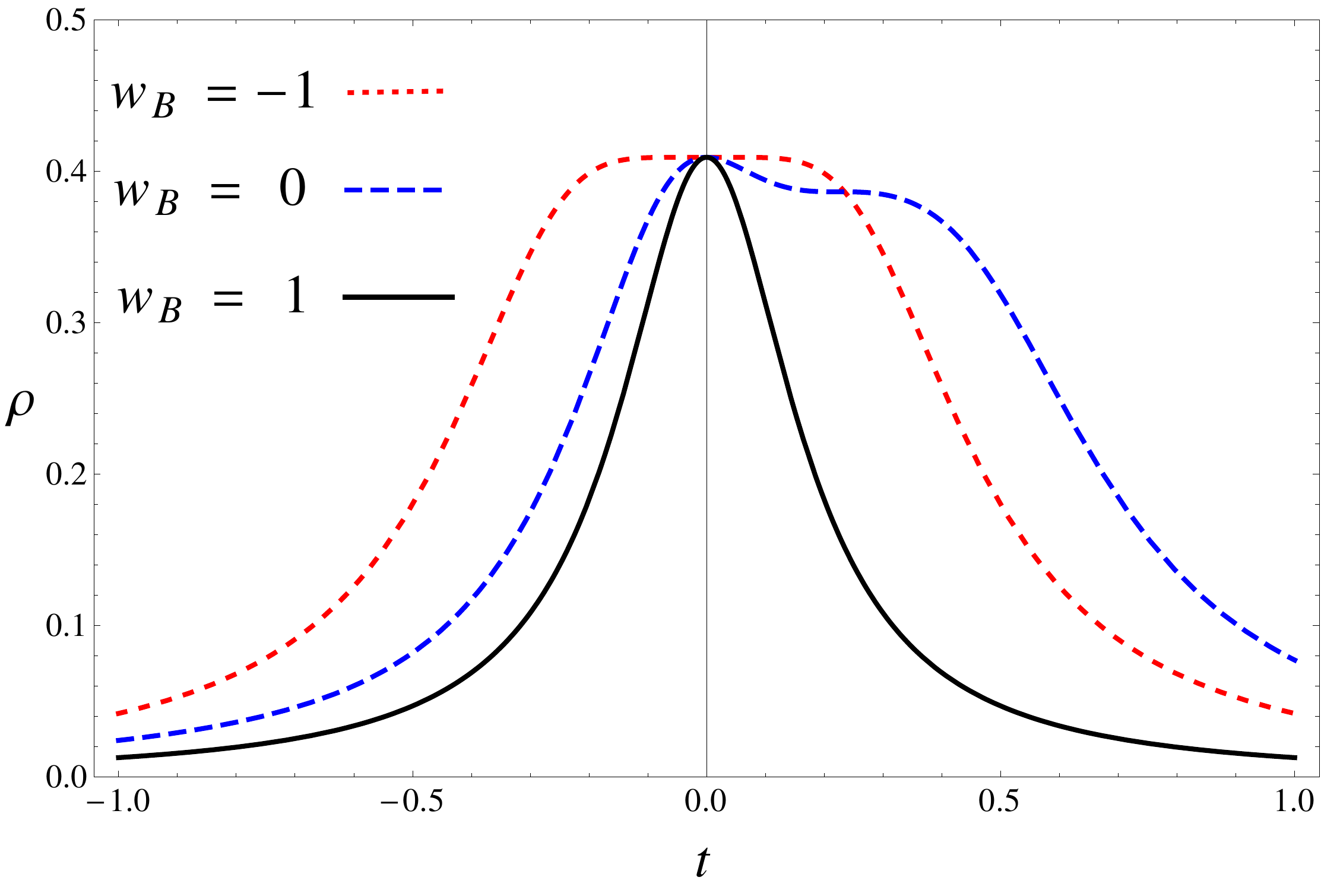} 
\caption{These plots show the Ricci scalar (left panel) and energy density (right panel) near the bounce for $\phidb<0$. Similar to \fref{fig:ricci}, the Ricci scalar at the bounce is zero for the extreme potential energy  dominated case ($w_{\rm B} = -1$) and maximum for the extreme kinetic energy dominated case ($w_{\rm B} = 1$). Also, the energy density is maximum at the bounce for all cases.
For the intermediate case when the potential and kinetic energy are equal at 
the bounce ($w_{\rm B} = 0$), the maximum of the Ricci scalar does not occur at the 
bounce. For all initial conditions with $w_{\rm B}<1$, the Ricci scalar behaves non-monotonically in the future evolution. Moreover, the bounce is time-reversal symmetric only for $w_{\rm B}=1$ and $w_{\rm B}=-1$. It is 
noteworthy that for $-1<w_{\rm B}<1$, the behavior of $\mathcal R$ and $\rho$ is the 
mirror image of those in \fref{fig:ricci}.}
\label{fig:riccineg}
\efig

\subsubsection{Negative inflaton velocity: $\phidb<0$}
Let us now consider initial data with negative initial inflaton velocity.
\fref{fig:riccineg} shows the evolution of the Ricci scalar and the energy
density for three different equations of state at the bounce: $w_{\rm B}=1,0,-1$. 
Similar to the
positive $\phidb$ case, $\mathcal R$ at the bounce is zero for $w_{\rm B}=-1$,
maximum for $w_{\rm B}=1$ and half the maximum value for $w_{\rm B}=0$. Also, the evolution of $\mathcal R$ is time-reversal symmetric for $w_{\rm B}=1$ and $w_{\rm B}=-1$, but asymmetry for all other initial conditions. Interestingly, this time reversal asymmetry is the mirror image of the asymmetry for positive $\phidb$ (\fref{fig:ricci}). 

Similar to \tref{tab:posphid}, \tref{tab:negphid} shows some of the representative simulations for various initial conditions. The first row
corresponds to $w_{\rm B}=0$ and all other rows have $w_{\rm B} \approx 1$. It turns out that, similar to the positive $\phidb$ case, the potential energy  dominated initial conditions
do not lead to the desired slow-roll phase. The second row corresponds to the extreme kinetic energy dominated case with exactly zero potential energy at the bounce. 
Unlike in the the positive $\phidb$ case this initial condition does not contain the desired slow-roll phase in its future evolution. The reason for this asymmetry between  positive and negative $\phidb$ is that for negative $\phidb$ the inflaton
initially evolves to the left of the potential (see \fref{fig:pot}) and in fact reaches $\phi_*$ very quickly, however, when it does so, its kinetic energy is too large and therefore the slow-roll parameters are too large and the desired phase of slow-roll does not start. This does not immediately imply that slow-roll will not occur at all, because when the field rolls back from the left side of the potential it might be possible that it climbs up the potential and reach $\phi_*$ again, now with significantly less kinetic energy. Simulations show that this scenario did not happen, because the field  
loses a lot of kinetic energy due to both Hubble friction and the steepness of the potential, consequently, it does not have enough kinetic energy to climb up the potential on the slow-roll side and reach $\phi_*$. 
Thus, this means that for the desired slow-roll phase to happen with $\phidb<0$, the inflaton field has to start high enough on the slow-roll side. 
As the table shows, $\phib \geq 3.63$ for the desired slow-roll phase to occur in the future evolution.

\bfig
\ig[width=0.7\textwidth]{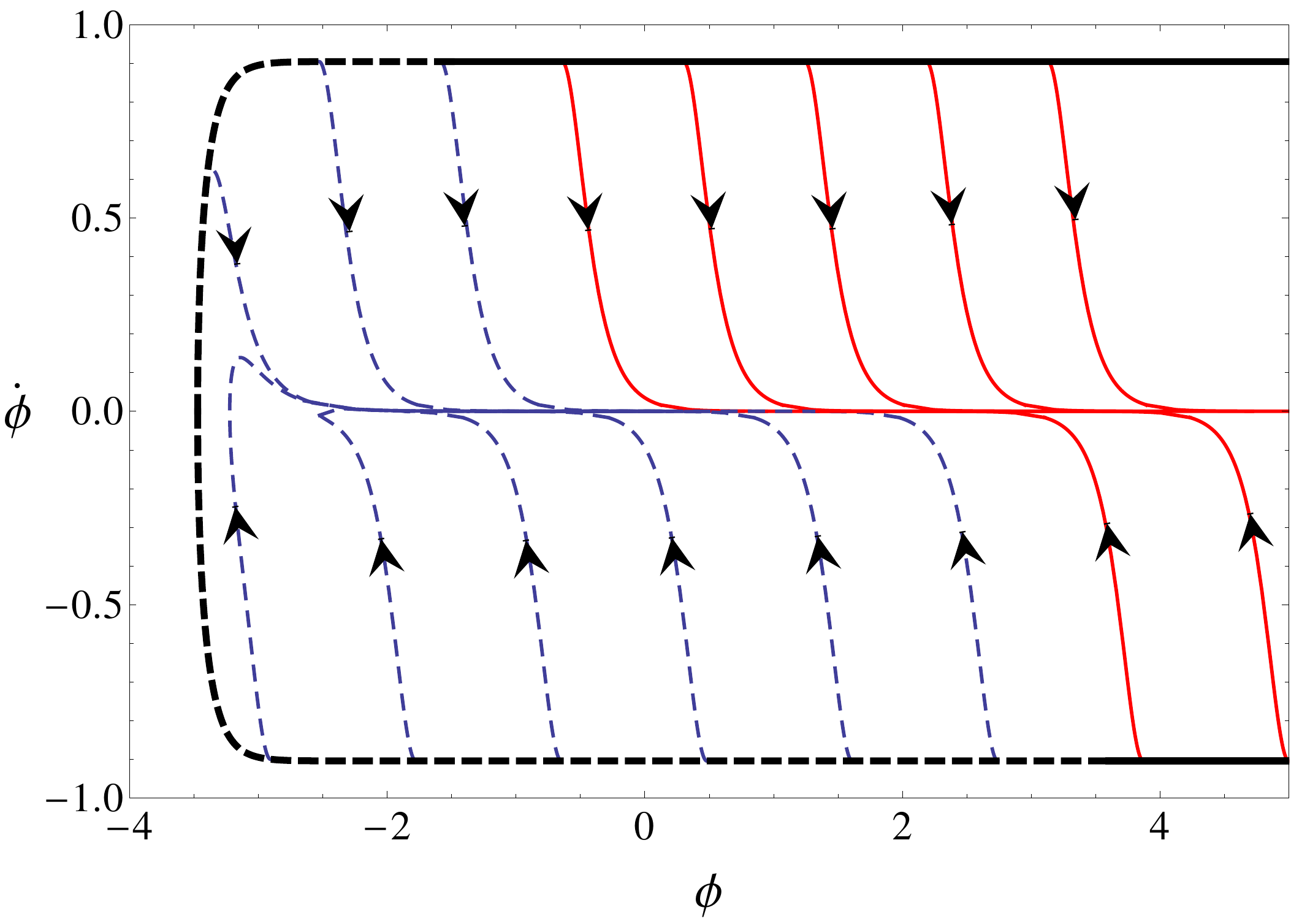}
\caption{$\phi-\dot\phi$ phase portrait of evolution trajectories for Starobinsky
potential in LQC. All trajectories (shown by curves with arrowheads) start at 
the bounce which is given by constant density surface $\rho=\rhomax$ (shown by the 
thick black boundary curve without arrowheads). The bounce surface extends from 
$\phi=-3.47~\mpl$ all the way to $\infty$, but here we have only shown a part of
it. The dashed (blue) curves do not lead to the desired slow-roll, while the 
solid (red) ones do. Since the diagram extends all the way to $\phi=\infty$, 
the fraction of dashed (blue) curves is extremely small compared to the 
solid ones. Therefore, it can be concluded that the occurrence of slow-roll is 
almost inevitable.}
\label{fig:phase}
\efig

\subsubsection{Phase portrait and the desired slow-roll}
Looking at \tref{tab:posphid} and \tref{tab:negphid}, we conclude that
observationally compatible initial conditions are: $\phib\geq-1.45 \, \mpl$ for
positive $\phidb$ and $\phib\geq3.63  \, \mpl$ for negative $\phidb$.
{\it Thus, in the entire parameter space of the initial conditions it is only 
the kinetic energy dominated initial conditions which lead to the desired 
slow-roll phase in their future evolution; there is a subset, rather small 
however, of the kinetic energy dominated initial conditions which do not give 
desired slow-roll.} \fref{fig:phase} shows some of the representative 
trajectories of evolution in a $(\phi, \, \dot\phi)$ phase diagram starting 
from the initial data surface (thick black curve without arrowheads). As 
discussed in \sref{sec:initial} the initial data surface is non-compact: while 
$|\phidb| < 0.91 \,\mpl^2$, $\phib$ is not restricted to a finite interval, 
that is, $\phib \in \left[-3.47 \, \mpl, \infty \right)$.  The blue, dashed 
trajectories are the ones which do not lead to the desired slow-roll in their 
future evolution, while the red, solid trajectories are the ones which do. In 
the same way, the dashed part of the initial data surface corresponds to the 
subset of initial data that is not compatible with observations while the solid 
part (which continues to $\phi\rightarrow\infty$) is compatible. Clearly, the 
part that does not lead to the desired phase of slow-roll is negligible 
compared to the entire initial data surface. In this sense, a significantly 
high fraction of initial data lead to the desired slow-roll and inflation is 
almost inevitable.

\subsubsection{Comparison with quadratic potential}
There are some interesting differences and similarities between inflation with a quadratic potential and with the Starobinsky potential. In particular, while potential  energy dominated bounces lead to an enormous amount of inflation for the quadratic potential, they do not lead to enough inflation to be compatible with observations for the Starobinsky potential. As a result, in Starobinsky inflation, only (a large subset of) kinetic energy dominated bounces are compatible with observations, whereas for inflation with a quadratic potential both kinetic and potential energy  dominated bounces are compatible with observations. 

For kinetic energy dominated bounces, the details of the potential do not matter in the early evolution, in other words, the evolution is driven by the kinetic energy. Consequently, the evolution of a kinetic energy dominated bounce in Starobinsky inflation will be similar to the evolution of a kinetic energy dominated bounce with a quadratic potential. Since the quantum gravity effects stem from exactly this period, one can expect that the quantum geometry corrections to the standard inflationary paradigm will be similar for Starobinsky inflation and kinetic energy dominated bounces in a quadratic potential. This expectation will indeed be borne out in the next subsections. There are, however, some surprises for extremely long wavelength modes that are larger than the observable universe. 

To see explicitly that the evolution is indeed similar, \fref{fig:comparison} shows a plot of the scale factor, which captures all the geometric information, evolved for inflation with a scalar field in the quadratic potential and Starobinsky potential. In order for this comparison to be meaningful, we took $w_{\rm B}$ to be the same for both models, in particular, $w_{\rm B}=0.9999999618$. This does not completely fix $\phib$ and $\phidb$: there is some freedom in overall signs left. We took the signs to be the same for both models, specifically, $\phib=-92.14 \, \mpl$ and $\phidb>0$ for the quadratic potential and $\phib=-1.45 \, \mpl$ and $\phidb>0$ for the Starobinsky potential. Note that around $t \approx 600 \, \spl$ the two geometries start to deviate slowly. This is when $w \approx 0.6$ for the quadratic potential (for the Starobinsky potential it is still very close to its value at the bounce). By the time $t=1168 \, \spl$, the scalar field in the quadratic potential has lost half of its initial energy to potential energy and the two geometries start to deviate more and more. Thus, indeed the details of the potential are irrelevant close to the bounce for kinetic energy dominated bounces as far as the background geometry is concerned. Note that this is true for \textit{all kinetic energy dominated bounces and does not require the initial conditions to be `close' to the bottom of the potential}, where the Starobinsky potential can be considered to be well approximated by the quadratic potential. Clearly, after the kinetic energy equals the potential energy, the features of the potential become more important. This occurs, however, well into the regime where classical GR is in excellent agreement with the effective equations of LQC and is thus irrelevant for the quantum geometric corrections (but \textit{is} very important for the predictions within the standard inflationary paradigm!).

%\todo{uncomment figure (and its caption) below once the figure is done :)}
\bfig
\ig[width=0.7\textwidth]{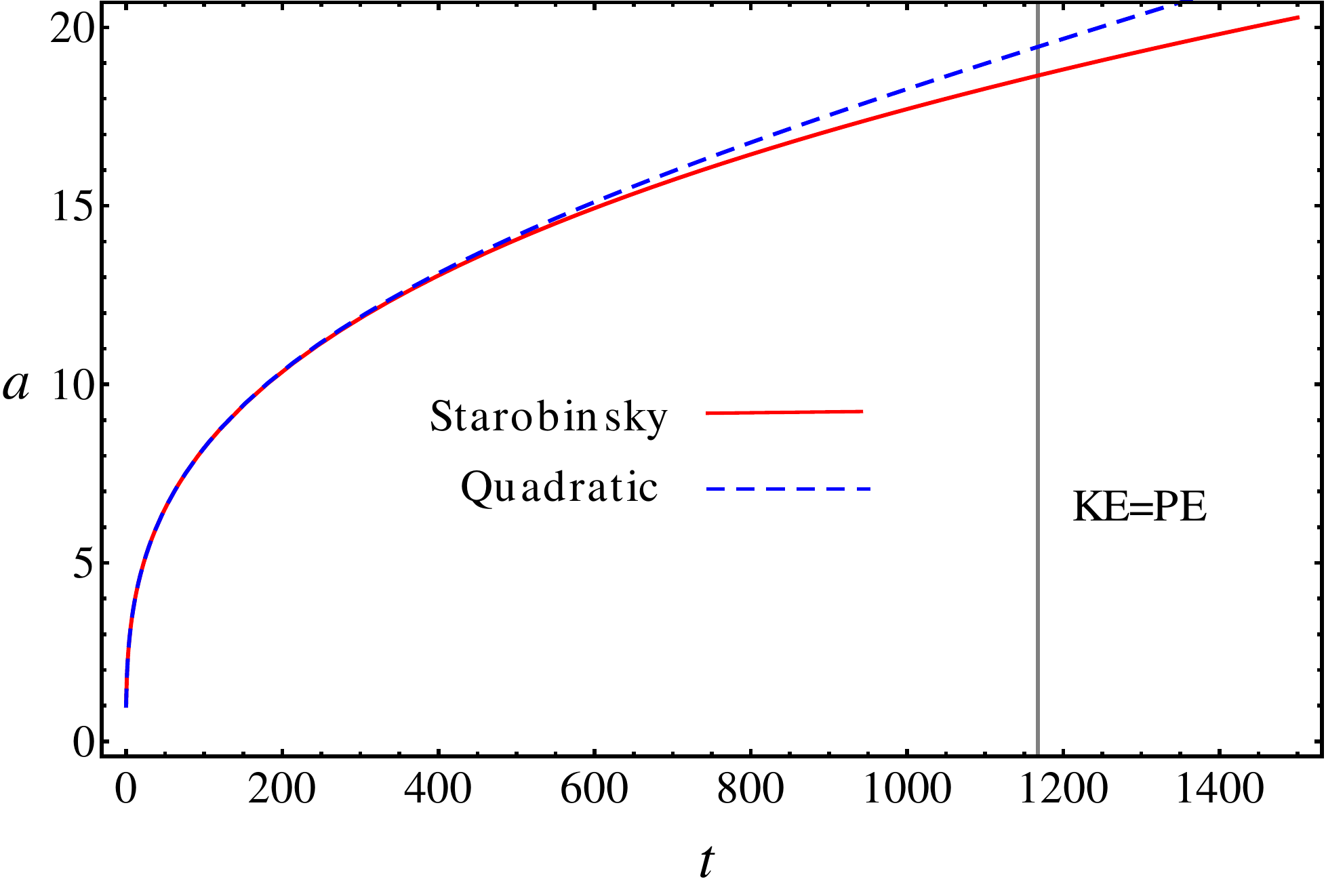}
\caption{Plot of the evolution of the scale factor for two different cases: 
quadratic potential (blue, dashed) and the Starobinsky potential (red, solid). 
The gray vertical line at $t=1168 \, \spl$ indicates when the kinetic energy 
becomes equal to the potential energy for the quadratic potential (for the 
Starobinsky potential this occurs much later, see \tref{tab:negphid}). Both 
cases are kinetic energy dominated bounces with the same value of $w$ at the bounce, 
that is, $w_{\rm B}=0.9999999618$. This determines $\phib$ and $\phidb$ up to 
signs. We choose the same signs for the quadratic and Starobinsky potential for 
easier comparison. In particular, $\phib=-92.14 \, \mpl$ and $\phidb>0$ for the 
$\phi^2$ potential and $\phib=-1.45 \, \mpl$ and $\phidb>0$ for the Starobinsky 
potential.}
\label{fig:comparison}
\efig

\subsection{Phenomenological considerations}
\label{sec:phenom}
\bfig
   \includegraphics[width=0.45\textwidth]{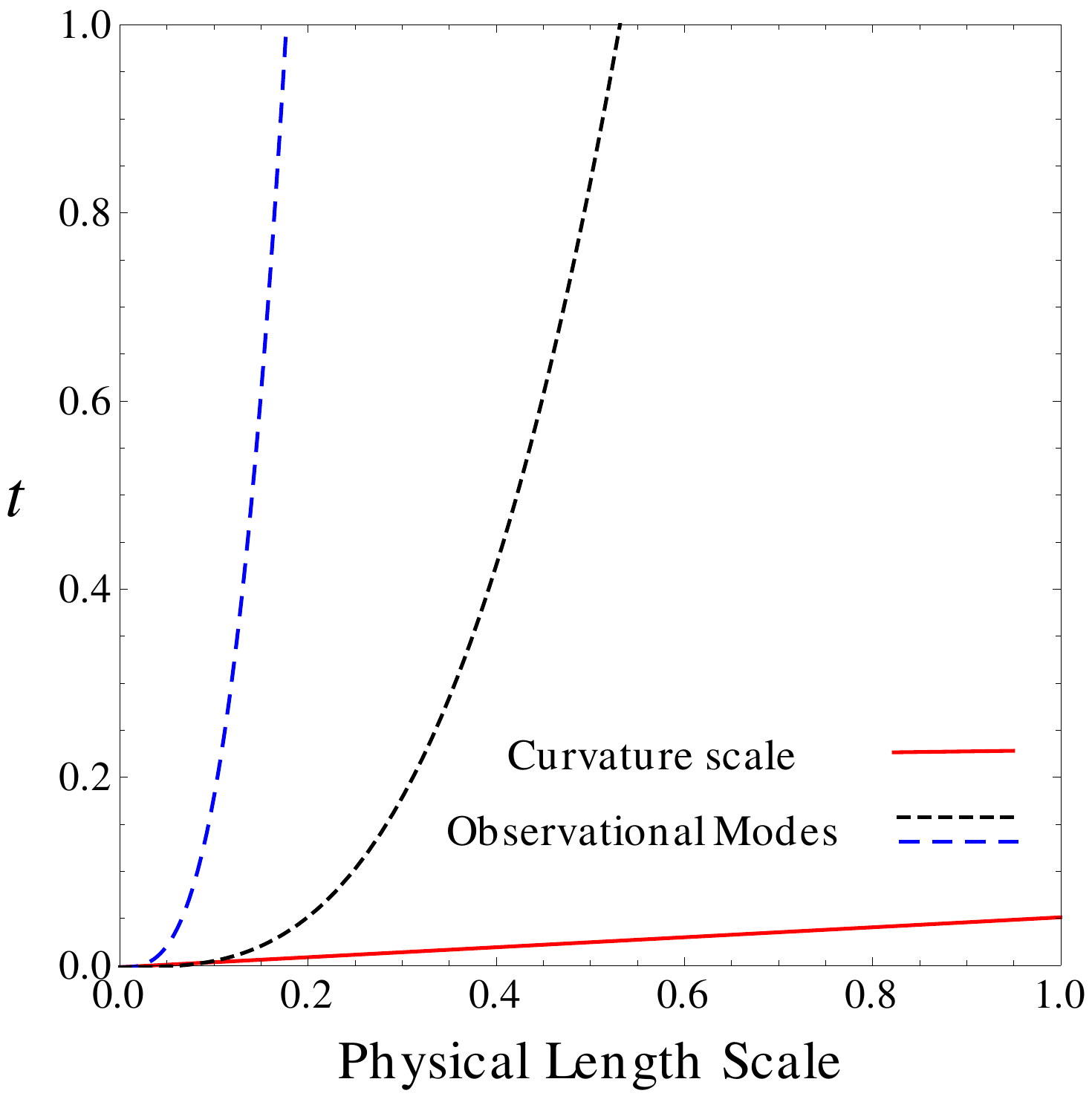}
\hskip0.5cm
   \includegraphics[width=0.45\textwidth]{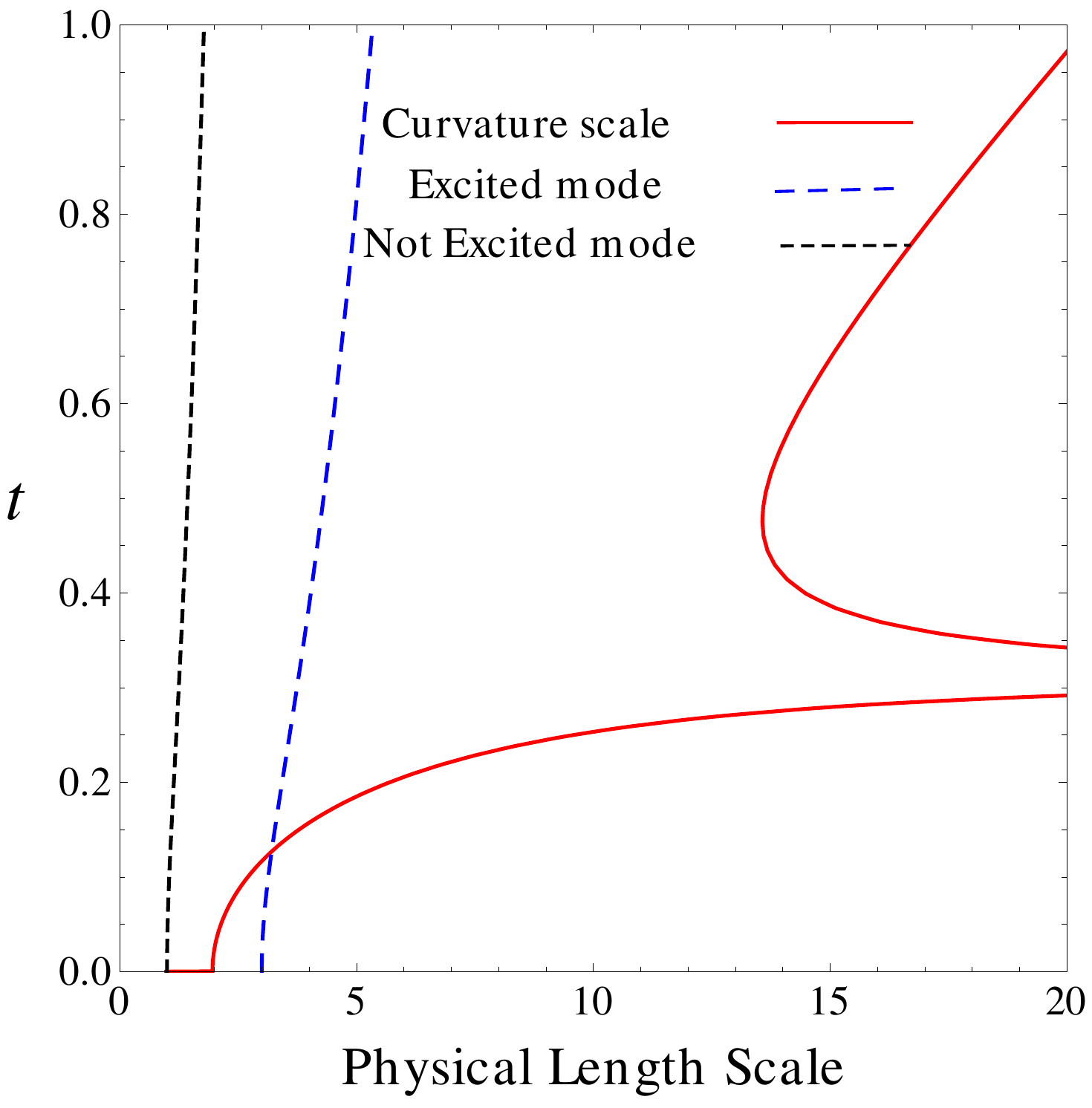}
\caption{Evolution of two observable modes and the physical curvature length 
scale: in the classical theory close to singularity (shown in the left panel) 
and in LQC close to the quantum bounce (shown in the right panel). It is clear 
that the physical wavelengths of both modes remain smaller than the curvature 
length scale in the classical theory all the way to the big bang. Whereas one of 
the modes in LQC becomes larger than the non-zero curvature length scale at the 
bounce. This mode gets excited in the future evolution.}
\label{fig:lqcgr}
\efig
Before we move to the discussion of the power spectra of quantum perturbations, 
some phenomenological considerations are in order. So far, in this paper, we 
have seen that the inflationary scenario with a Starobinsky potential 
(\ref{eq:pot}) fits well with the non-singular bouncing picture of LQC: The 
pre-inflationary dynamics of the background spacetime is significantly modified 
due to the quantum geometric effects for energy scales in the regime $10^{-3} 
\rhomax < \rho < \rhomax$. As a result, quantum perturbations in LQC experience 
a very different history before the onset of inflation compared to the quantum 
perturbations in the standard inflationary scenario based on GR. 
%For instance, the curvature scalars in LQC are finite throughout the evolution and become maximum at the bounce. 
There is a finite non-zero length scale associated to 
the  maximum value of the Ricci curvature scalar, 
$k_{\rm LQC}:=(\mathcal R_{\rm B}/6)^{1/2}$, due to which modes of different 
wavelengths interact with the curvature in a qualitatively different way.
Ultraviolet modes ($k\gg k_{\rm LQC}$) are too energetic to be affected by 
the curvature and 
evolve as if they are in a flat spacetime, whereas mode with small wavenumber 
($k\sim k_{\rm LQC}$) do get excited by the curvature. This can be understood 
via \fref{fig:lqcgr}, which shows the evolution of two observable modes in GR 
(in the left panel) and LQC (in the right panel). In GR, all observable modes 
remain smaller than the curvature length scale all the way to the big bang. In 
LQC, on the other hand, there are some modes which are larger than the 
curvature length scale near the bounce. It is these infrared modes that will 
be excited in their future evolution and deviate from the usual Bunch-Davies 
state at the onset of inflation.

CMB experiments, such as Planck and WMAP, report the value of the amplitude 
and the spectral index of the power spectrum at a reference mode which is
given by a comoving wavenumber $k_*$ whose physical wavenumber 
($k_*/a(t_*)$) at the time of the horizon crossing is equal to the 
Hubble parameter at that time, i.e. $k_*/a(t_*)=H_*$. Since $k_*$ is 
the comoving wavenumber, its numerical value depends on the numerical value of 
the scale factor which is different for different conventions. For example, in 
standard cosmology and also in the phenomenology of CMB experiments the scale 
factor is chosen to be unity today, whereas in our numerical simulations we 
have chosen the scale factor to be unity at the bounce. Therefore the numerical 
value of the comoving wavenumber of the reference mode in LQC, $\ksLQC$,
will be different from the comoving wavenumber $k_*$ at which $\as$ and $\ns$ 
are reported. However, the physical wavenumber corresponding to both $\ksLQC$ 
and $k_*$ are still the same, i.e. $\ksLQC/a_{\rm LQC}(t_*)=k_*/a_{\rm BD}(t_*)$.
Thus in order to make contact with the observational data we need to find the 
numerical value of $\ksLQC$ that has the same physical wavelength as the 
reference mode $k_*$ at the time of horizon crossing. We adopt the following 
strategy. We compare the LQC power spectrum at the end of inflation with those 
predicted by observation and then numerically search for $k$ at which the LQC 
power spectrum has the same amplitude and Hubble parameter as reported by 
observations. This $k$ is $\ksLQC$. Naturally, the value of $\ksLQC$ depends 
on the amount of e-folds between the onset of slow-roll and the bounce, which 
can be related to the initial conditions at the bounce: $\phib$ and the sign 
of $\phidb$. This relation is shown in \fref{fig:kstar}.

\bfig
\ig[width=0.48\textwidth]{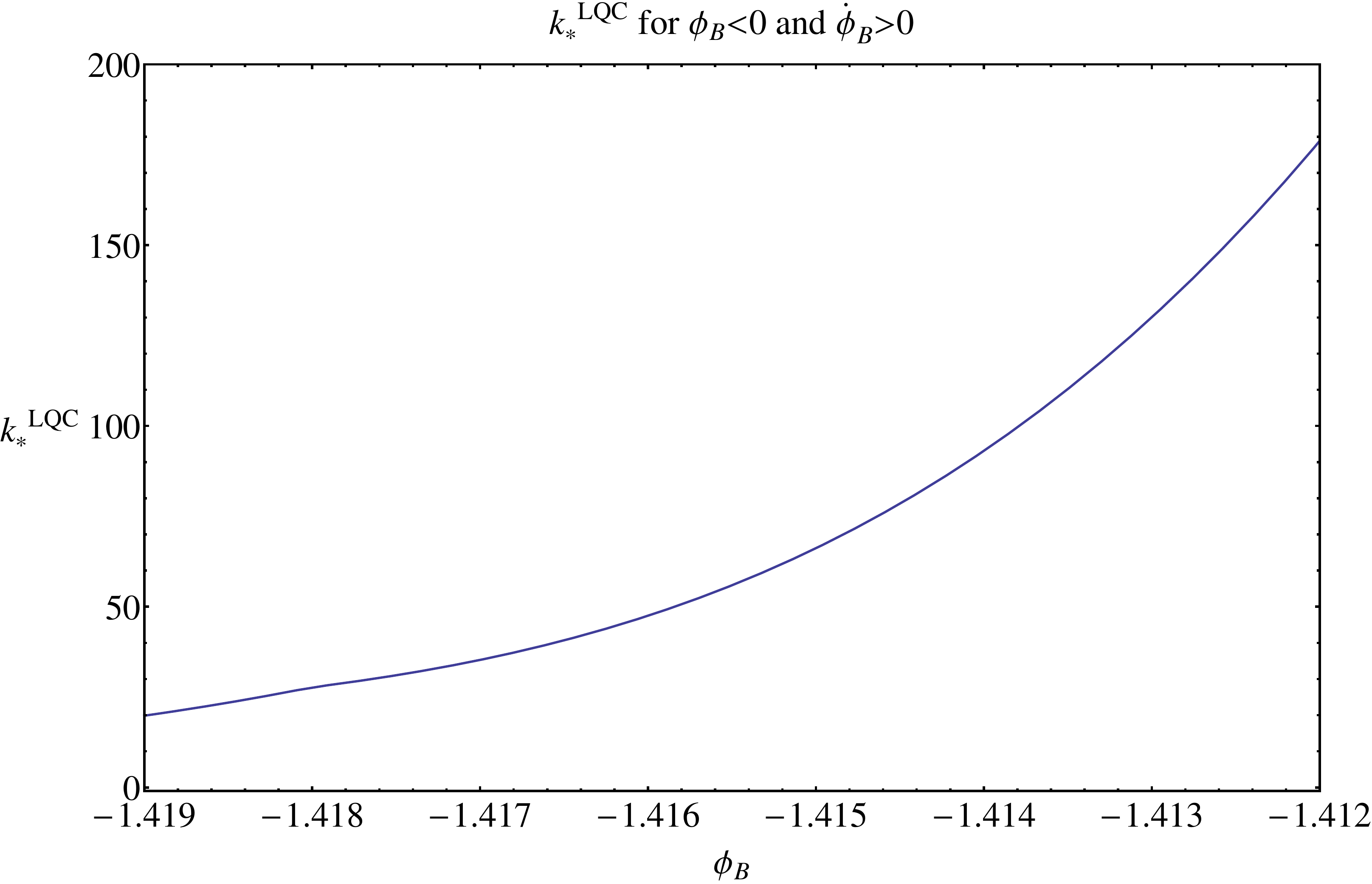}
\hskip0.4cm
\ig[width=0.48\textwidth]{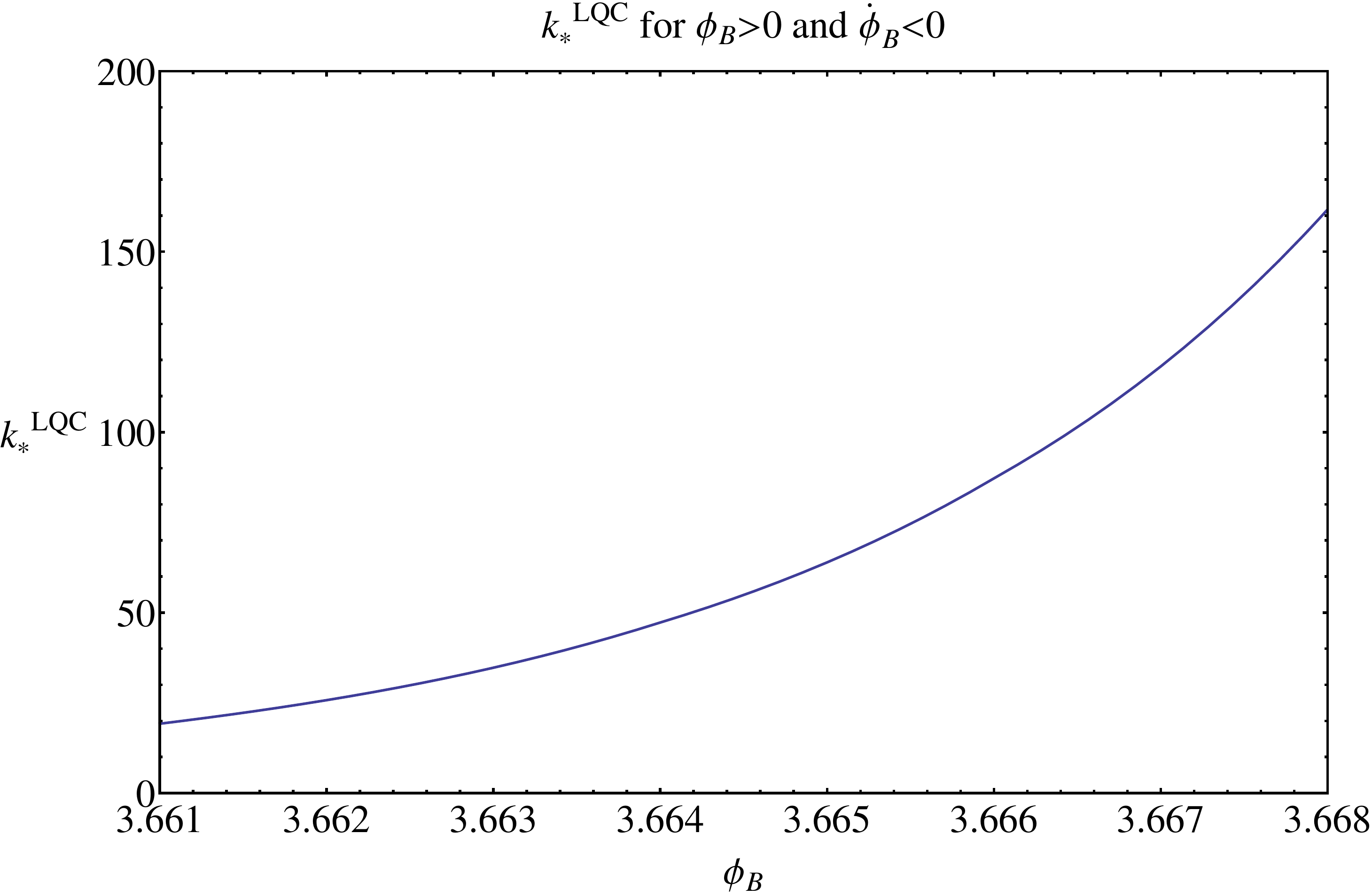}
\caption{The reference mode $k^*$ shown as a function of $\phib$ which controls
the number of e-folds. It is apparent that $k^*$ increases with increasing 
$\phib$.}
\label{fig:kstar}
\efig

As inflation happens, the infrared modes get stretched to super Hubble size and 
may no longer fall into the observable window. Therefore not all modes of the
perturbations are observable, rather, there is a finite range: $(k_{\rm min}, 
k_{\rm max})\approx(\ksLQC/8.58,\,200 \ksLQC)$, where $(k_{\rm min}$ and $k_{\rm max}$ 
are the co-moving wavenumbers of the longest wavelength entering the Hubble horizon today 
and the shortest observable modes in the CMB, respectively.\footnote{
$k_{\rm min}=a_o~H_o$, where $a_o$ is the scale factor today and 
$H_o=0.000233~Mpc^{-1}$ is the Hubble constant \cite{wmap7}. The pivot scale is taken 
to be $\ks/a_o=0.002~Mpc^{-1}$. Using these relations: 
$k_{\rm min}\approx\ks/8.58$, and $k_{\rm max}\approx2000 k_{\rm min}\approx 200 \ks.$}
If $\phib$ is changed, the 
value of $\ksLQC$ changes as shown in \fref{fig:kstar}, which will shift the
observational window in which LQC effects are observable. On the one 
hand, if $\ksLQC$ is too large then LQC corrections may not be observable at 
all. On the other hand, if $\ksLQC$ is too small then the deviations between 
the standard inflationary power spectrum and LQC power spectrum may be too 
large to agree with current observations. Additionally, if $\ksLQC$ is
too small, numerical investigations have shown that the test field 
approximation may fail. By demanding that the LQC power spectrum agrees with 
the standard power spectrum for $\ell \gtrsim 30$ and the backreaction of the
perturbations during quantum gravity regime is small, we can further restrict 
the set of initial conditions that are both observationally interesting and 
self-consistent with the framework of quantum field theory on quantum 
spacetime. For a type-I state whose power spectrum is shown in \fref{fig:power},
above requirements lead to the following observationally relevant initial
conditions: $-1.419~\mpl < \phib < -1.412~\mpl$ for positive $\phidb$ 
and $3.661~\mpl < \phib< 3.668~\mpl$ for negative $\phidb$ at the bounce. Note
that these initial conditions might vary for different states that have
different power spectra.

\subsection{Perturbations}
\label{sec:resultpert}
As discussed earlier in this section, depending on the initial value of the 
inflaton field, the bounce and its subsequent quantum gravity regime can be 
potential or kinetic energy dominated. For the kinetic energy dominated bounces 
(with $w_{\rm B}\approx 1$), the background evolution in the quantum gravity 
regime is practically the same for all potentials, because the scalar field 
behaves like a massless scalar field. For these initial conditions, the quantum 
gravity corrections to the inflationary power spectrum for the Starobinsky 
potential turn out to be qualitatively very similar to that for the quadratic 
potential. For potential energy  dominated bounces, on the other hand, the 
quantum gravity regime is dictated by the shape of the potential and hence one 
would expect potential specific features in the inflationary power spectrum. 
However, as we found in the first part of this section, potential energy  
dominated initial conditions are not compatible with observations for 
Starobinsky inflation as they do not lead to the desired slow-roll phase. 
Hence, given initial conditions for a quadratic potential and Starobinsky 
potential that lead to the same number of e-folds from the bounce till the
end of inflation with $N\gtrsim60$, the LQC corrections to the observable 
inflationary power spectra in the two cases are practically indistinguishable. 
In the following, we will discuss the scalar and tensor power spectra. The 
initial conditions for perturbations are given slightly before the bounce and 
then evolved till the end of inflation using \eref{eq:scalar} and (\ref{eq:tensor}). 

%Scalar power spectrum without kstar
\bfig
\ig[width=0.70\textwidth]{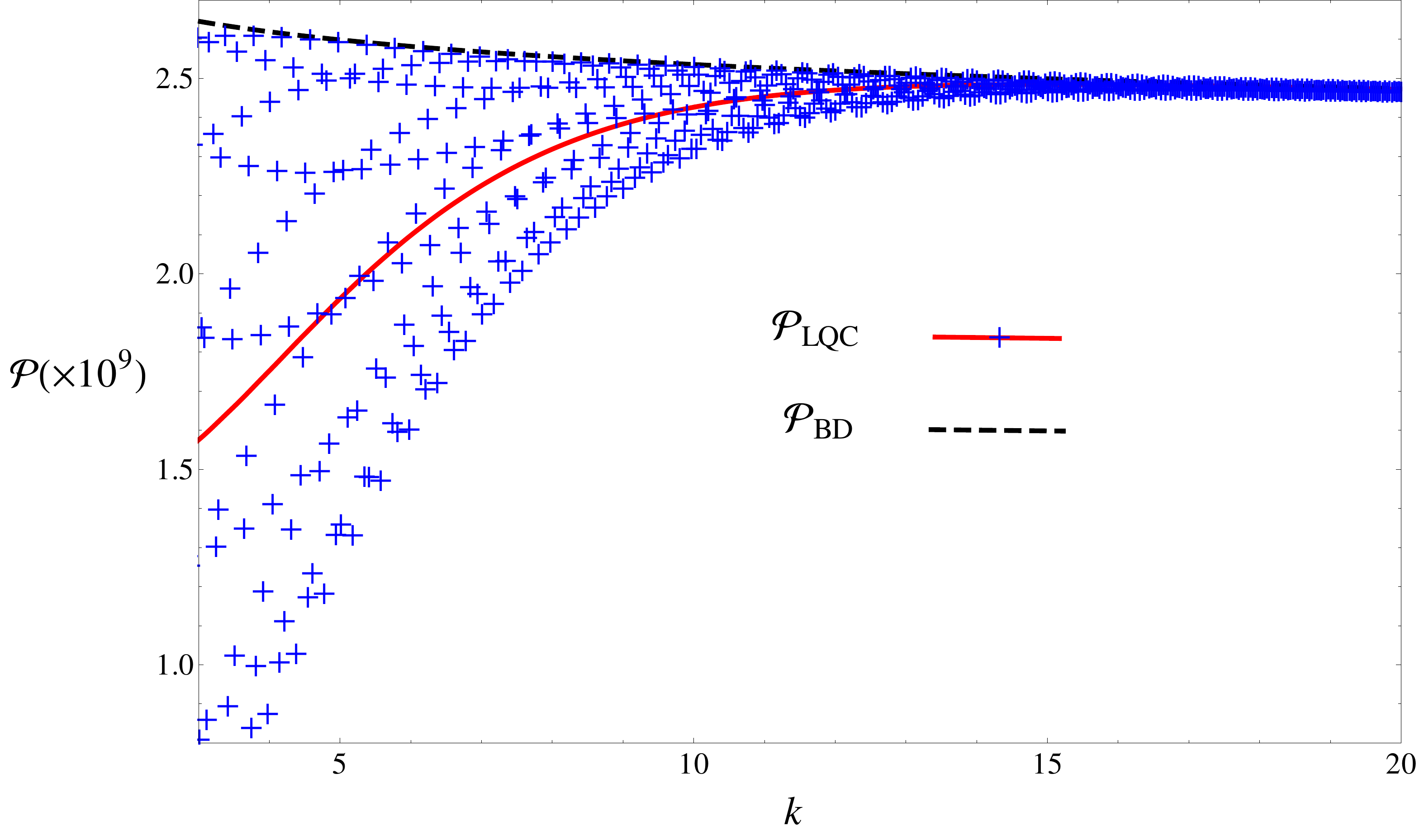} 
%\hskip0.2cm
%\ig[width=0.47\textwidth]{figs/ns.pdf} 
\caption{Scalar power spectrum (left panel) for $\phib=-1.419 \, \mpl$ 
for Starobinsky potential using the type I initial state. The red, solid curve shows
the binned LQC 
spectrum, the black, dashed curve shows the Bunch-Davies power spectrum 
without LQC modifications and the blue scattered points show the un-binned 
LQC power spectrum.  It is apparent that for large $k$ the LQC power spectrum 
agrees extremely well with the standard inflationary power spectrum. This 
behavior is similar to that of a quadratic potential as shown in \cite{aan3}.
}
\label{fig:power}
\efig

\subsubsection{Scalar modes}
\fref{fig:power} shows the scalar power spectrum for a representative case with
$\phib=-1.45 \, \mpl$ and $\dot\phi_{\rm B} > 0$. This corresponds to a total 
of $\sim67$ e-folds from the bounce till the end of inflation. Similar to the 
quadratic potential (\cite{aan3}), the true LQC power spectrum is oscillatory 
as shown by the blue ``$+$'' points in the figure. This oscillatory behavior 
of the power spectrum can be understood by first writing the mode function of 
the scalar perturbations as Bogoliubov transformation on the usual Bunch-Davies
states:
\be
  q_{\rm LQC}(k) = \alpha (k)~q_{\rm BD}(k) + 
                        \beta^*(k)~q^*_{\rm BD}(k).
\ee
The LQC power spectrum then can be written as:
\be
\plqc^{(s)}(k) = \pbd^{(s)}(k) |\alpha(k)+\beta^*(k)|^2 
         = \pbd^{(s)}(k)\left(|\alpha(k)|^2+|\beta(k)|^2+2 |\alpha(k)||\beta(k)|
\cos(\theta(k))\right),
\ee
where $\theta$ is the relative phase between $\alpha(k)$ and $\beta(k)$
and the superscript `$(s)$' denotes scalar modes.
It is the cosine term in the above expression that is responsible for the 
oscillatory behavior of the power spectrum. This term vanishes when averaged 
over a window of $k$, and only the $|\alpha(k)|^2+|\beta(k)|^2$ 
part remains. It turns out that these oscillations are too rapid to be seen in 
the CMB as they will be averaged out by the spherical Bessel functions while 
computing $\mathcal C_{\ell}$ at the surface of last scattering.\footnote{
$\mathcal C_{\ell}^{TT}=\int_k \rm (d\ln(k)) \mathcal P(k) \left[ j_{\ell}(k) 
\Theta(k)\right]^2$, where $\mathcal P$ is the power spectrum, $j_{\ell}$ is 
the spherical Bessel function, and $\Theta$ is the transfer function.} 
Therefore, what matters for observations is the average value of the power 
spectrum: $|\alpha(k)|^2+|\beta(k)|^2$, which can also be obtained 
by binning the oscillatory power spectrum. 
%The corresponding spectral index $\ns:=d\plqc/d\ln(k)$ is shown in \fref{fig:ns}. 
%As expected $\ns$ remains the same as that for BD states for large $k$, whereas there 
%are deviations for small $k$.
%\bfig
%\ig[width=0.50\textwidth]{figs/Power_ks.pdf} 
%\ig[width=0.65\textwidth]{figs/ns_k.pdf} 
%\caption{Spectral index of the scalar power spectrum $\ns:=d\plqc/d\ln(k)$ for
%$\phib=-1.417 \, \mpl$ and $\phidb>0$. The red, solid curve
%shows $\ns$ of the averaged scalar power spectrum (red solid curve in
%\fref{fig:power}) and the dashed curve shows the spectral index of the BD power
%spectrum as reported in \cite{planck15xx}. It is apparent that the two curves 
%deviate for small $k$ while there is great agreement for large $k$. This is
%because only small $k$ modes are excited during the evolution.}
%\label{fig:ns}
%\efig
Therefore, as far as observations are concerned the LQC corrections to the 
standard power spectrum is simply a factor of 
$|\alpha(k)|^2+|\beta(k)|^2=1+2 |\beta(k)|^2$ 
(shown by the red, solid curve in \fref{fig:power}), where $|\beta(k)|^2$ is the 
number density of the particles produced by the pre-inflationary dynamics of 
LQC with respect to the standard Bunch-Davies vacuum. 

The figure clearly shows that only the small $k$ modes deviate from the 
standard Bunch-Davies power spectrum, while for large $k$ there is remarkable 
agreement with the standard Bunch-Davies power spectrum. As discussed before, 
this behavior stems from the fact that modes with $k$ smaller than the 
characteristic curvature scale ($k_{\rm LQC}= (\mathcal R_{\rm B}/6)^{1/2}$) are 
excited in the quantum gravity regime (\sref{sec:initial}) and the particle 
density for those modes is non-zero ($|\beta(k)| > 0$). For larger wavenumbers,
on the other hand, $|\beta(k)|$ rapidly decays to zero because high $k$ modes 
are too energetic to be affected by the curvature and as a result do not get 
excited. 

It is apparent from the discussion above that the qualitative features of the 
LQC corrections to the {\it observable} power spectrum for the Starobinsky 
potential are the same as those for the quadratic potential, and hence robust under 
the choice of potential. This is surprising given that the dynamics of the 
inflationary phase is drastically different for both potentials. 

%\fref{fig:ns} shows the scalar power spectrum as a function of $k/k^*$ for
%$\phib=-1.417$ for which $k^*=...$. \red{how is it different from before?} It is 
%apparent from the figure that for $k\geq k^*$ the Bunch-Davies and LQC power 
%spectrum are is great agreement, while the differences are more pronounced for 
%$k/k^*\lesssim 0.3$. The power spectrum at the end of inflation when evolved 
%till the surface of last scattering, the $k^*$ correspond approximately to 
%multipole $\ell \approx 30$. Therefore, the LQC power spectrum shown 
%in \fref{fig:powerks} will be in great agreement with the standard inflationary 
%prediction at $\ell\approx30$ but there will be deviations for $\ell\lesssim 10$.  

\bfig
\ig[width=0.7\textwidth]{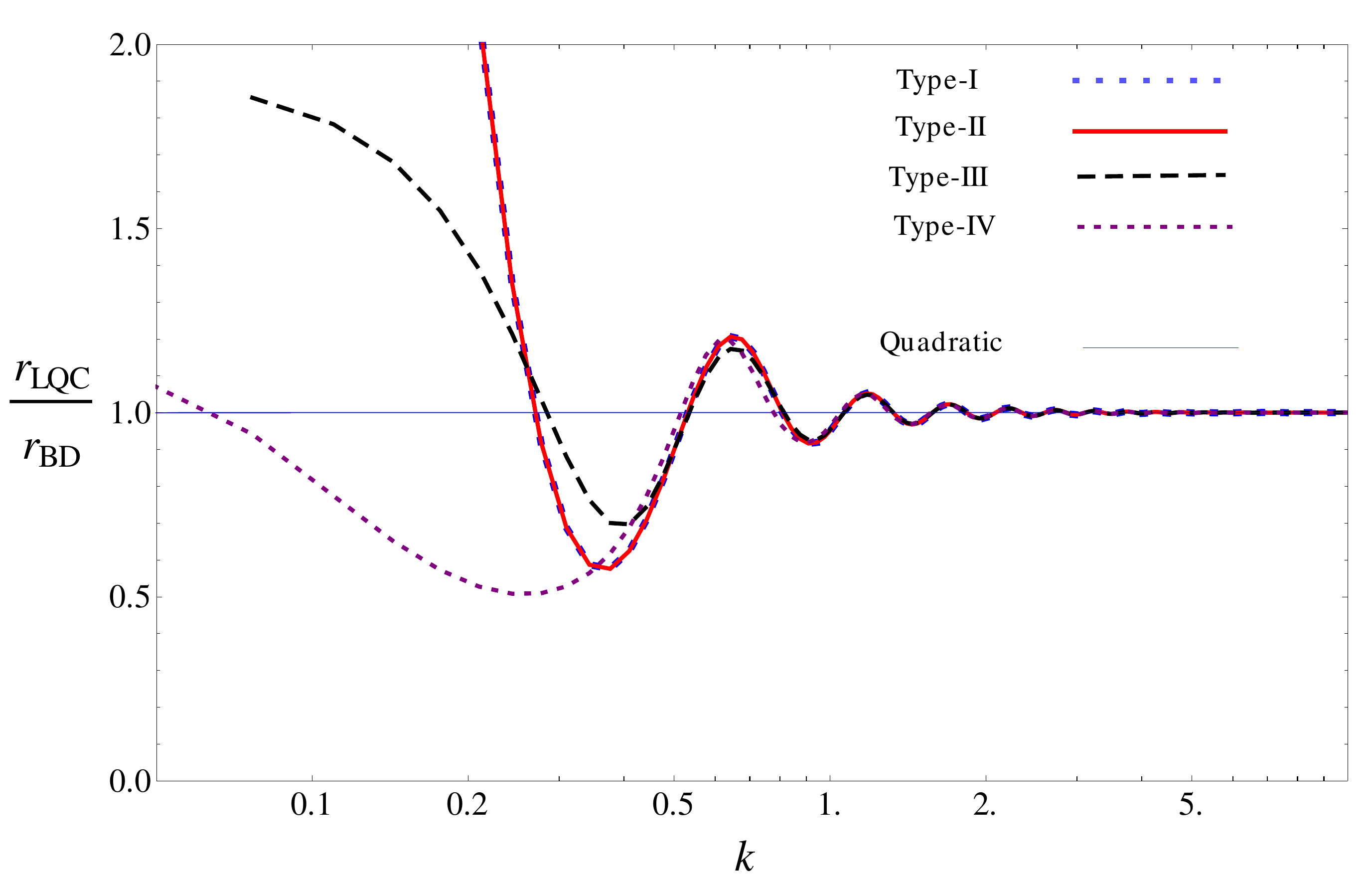}
\caption{Comparison of tensor to scalar ratios in LQC and standard inflation for
Starobinsky (with $\phib=-1.419 \, \mpl$ and $\phidb>0$) and quadratic 
potential (with $\phib=1.16~\mpl$). For the quadratic potential $r_{\rm LQC}$ 
remains extremely close to $r_{\rm BD}$ for all $k$ shown in the diagram while 
there are departures for the Starobinsky potential. These wavenumbers 
correspond to modes whose wavelengths are larger than the Hubble horizon. While 
this effect does not fall into the observable window, it can have non-trivial 
implications for tensor fossils and non-Gaussian modulation of the power 
spectrum.}
\label{fig:rlqcbd}
\efig
\subsubsection{Tensor modes}
Before moving to the discussion of tensor modes for the Starobinsky potenial let us
recall the results for the tensor spectrum with a quadratic potential 
obtained in \cite{aan3}. 
As apparent from \eref{eq:scalar} and (\ref{eq:tensor}), the scalar mode
evolution equation involves an effective scalar potential $\tilde 
{\mathcal U}$ which is absent for tensor modes. Therefore, in principle, the 
particle density for scalar and tensor modes should be different from each 
other for $k^2 \approx \tilde {\mathcal U}$. However, as shown in \cite{aan3}, 
the magnitude of $\tilde {\mathcal U}$ for the quadratic potential is of the 
order of $10^{-10}~k_{\rm min}^2$ (where, as before, $k_{\rm min}$ is the smallest value of 
$k$ that still falls in the observational range of $k$s); 
therefore the particle density for both the tensor and scalar modes is 
practically the same. Hence, a significant difference between the tensor and 
scalar perturbations would be apparent only for $k\sim 10^{-5} k_{\rm min}$ and 
since the wavenumber of the observable modes are much larger, this difference 
is not relevant for observations for the {\it quadratic potentials}. An immediate 
consequence of this for the quadratic potential is that the tensor to scalar 
ratio in LQC is the same as that in standard inflation, that is, 
$r_{\rm LQC}=r_{\rm BD}$ for all $k\gtrsim 10^{-5} k_{\rm min}$. 

While this is also true for most of the initial conditions for Starobinsky 
inflation, there does exist a small subset of the initial data surface 
($-1.45~\mpl\lesssim\phib\lesssim-1.38~\mpl$) for which $\tilde {\mathcal U}$ 
is of the order of $10^{-5} k_{\rm min}^2$. That is, for 
$k^2\gg k^2_{\rm min} 10^{-5}$ ($\Longleftrightarrow k\gg k_{\rm min}/300$), the effect of $\tilde {\mathcal U}$ 
on the scalar modes is negligible. 
As a result $|\beta(k)|$ is the same for scalar and tensor modes
and $r_{\rm LQC}=r_{\rm BD}$ for this range of $k$. 
On the other hand, for more infrared modes, i.e. 
$k \lesssim k_{\rm min}/300$, tensor and scalar particle densities are 
different and $r_{\rm LQC}\neq r_{\rm BD}$.  Thus, for some initial conditions, 
the effect of $\tilde {\mathcal U}$ shows up in Starobinsky inflation at a 
different scale than in the quadratic inflationary scenario. \fref{fig:rlqcbd} 
shows $r_{\rm LQC}/r_{BD}$ for the quadratic potential and the Starobinsky 
potential for various choices of initial vacua. It is apparent that, for 
the Starobinsky potential, $r_{\rm LQC}/r_{BD}$ shows a deviation from unity for infrared 
modes (small $k$), and becomes unity for large $k$ for all choices of initial 
vacua. For the quadratic potential, on the other hand, $r_{\rm LQC}/r_{BD}$ remains 
extremely close to unity for all $k$ and all choice of initial vacua. 
While these features are not relevant for observational modes, 
they could play an important role for 
three point functions and tensor fossils where one considers coupling of very 
long modes with observation modes. These effects could also lead to additional 
LQC signatures in non-Gaussian modulation of the power spectrum due to super 
horizon modes as considered in \cite{schmidthui,agulloassym}. This opens up new 
avenues to investigate various CMB anomalies from the perspective of quantum gravity.

%For Starobinsky potential, on the other hand, the magnitude of $\mathcal U$ in
%the quantum gravity regime is of the order of $10^{-6}$ which is a factor of
%1000 more than in the quadratic case. One would, therefore, expect departure
%from $r_{\rm LQC}=r_{\rm BD}$. The ratio of $r_{\rm LQC}$ to the standard 
%$r_{\rm BD}$ is shown in \fref{fig:rlqcbd}, where it is clear that there are
%small differences for $k\lesssim2$, however $r_{\rm LQC}/r_{\rm BD}$ quickly
%approaches to unity for $k>2$. If one considers more infrared modes with 
%$k~10^{-3}$, large deviations between $r_{\rm LQC}$ and $r_{\rm BD}$ can be
%found which may lead to running in the tensor spectral index for those modes. 
%But as before these modes are beyond the observation window so those
%effects may not of much interest. 

\section{Discussion}
\label{sec:disc}
Inflation is the most successful framework that generates appropriate 
initial conditions for the CMB which further seeds the formation of large 
scale structure observed today. There are, however, numerous 
inflationary models \cite{Martin:2013tda}. Thanks to remarkable advances in
CMB experiments over the past few decades many of these models have already 
been ruled out. For instance, recent data from the Planck mission show that the 
simplest inflationary model, that is, a single scalar field with a quadratic potential 
is moderately disfavored, while a scalar field with a Starobinsky potential is the 
most favorable of all \cite{planck15xx}. In the standard inflationary 
scenario the background spacetime is determined by classical GR which generically 
admits a big bang singularity in the past for all inflationary models \cite{borde}. 
Hence, the pre-inflationary dynamics of the spacetime remains elusive. 
For a satisfactory extension of the inflationary paradigm all the way to the 
Planck scale one needs a quantum theory of gravity. Primary challenges for such 
a theory would be: to resolve the classical singularity as well as admit an 
inflationary phase which is compatible with observations and occurs generically 
without requiring any fine tuning of the initial conditions. It is expected that 
such a fundamental theory would then also provide a consistent framework to 
study the evolution of cosmological perturbations.

This ambition is indeed achieved in LQC. Here, we performed a 
detailed numerical analysis of the quantum background spacetime and 
cosmological perturbations thereon for an inflationary model with a Starobinsky 
potential, which is favored by the data \cite{planck15xx}. 
A similar analysis was performed for the quadratic potential. 
There, it was shown that natural initial conditions exist at the bounce which 
give rise to a rich pre-inflationary phase--dominated by quantum gravity--that joins 
an observationally compatible inflationary phase quite generically \cite{asloan_prob2}. 
Due to quantum modified pre-inflationary dynamics, states describing the cosmological 
perturbations at the onset of inflation are different from the Bunch-Davies state 
and the resulting power spectra at the end of inflation acquire corrections compared to 
the standard power spectra \cite{aan1,aan3}. 
As discussed in \sref{sec:intro}, it is not a priori clear whether the pre-inflationary
dynamics and the LQC corrections obtained for the quadratic potential will also hold 
for the Starobinsky potential as these models differ significantly during inflation. 
We find that while most of the results concerning
LQC corrections to the power spectra obtained for the quadratic potential are also 
qualitatively true for the Starobinsky potential, there are some important 
differences. Our main results can be summarized as follows:
\begin{itemize}
\item {\bf Desired slow-roll almost inevitable:} As discussed in 
\sref{sec:results}, the range of initial conditions for the Starobinsky potential is 
semi-finite $\phib\in\left(-3.47,\infty\right)$. Only a small fraction--mostly
potential energy dominated--of the evolutionary trajectories starting from the 
initial data surface fail to give the desired slow-roll phase. Therefore, almost all 
initial conditions lead to inflation compatible with observation and the occurrence 
of the desired slow-roll phase is nearly inevitable {\it without requiring any fine 
tuning on the initial conditions}. 
Note that although this final conclusion is also true for the 
quadratic potential, where one found that the desired slow-roll phase occurred for nearly 
all initial conditions as well, its initial data surface at the bounce is
completely different from the surface for the Starobinsky potential. 
The range of initial conditions 
for the quadratic potential is finite: $|\phib| \in (0, \, 7.47 \times 10^5~\mpl)$. 
As a result of this difference, the way in which slow-roll is obtained is distinct 
for both potentials. 
For example, {\it none of the potential energy dominated initial conditions lead 
to occurrence of the desired slow-roll phase} in the Starobinsky potential as they do not
generate enough e-folds. Only kinetic energy dominated bounces lead to the desired 
slow-roll phase as opposed to the quadratic potential where all potential dominated 
conditions result in the desired slow-roll phase. 

\item {\bf Phenomenological robustness of observable LQC corrections:} 
LQC corrections to the power spectra of cosmological perturbations are born in 
the quantum gravity regime close to the bounce. Since only the kinetic energy 
dominated bounces for the Starobinsky potential are compatible with observations, 
the scalar field behaves like a massless scalar field (and thus the features
of the potential are not important) in the quantum gravity regime for the 
observationally relevant initial conditions. 
Interestingly, for the quadratic potential, it is exactly (a small fraction of) the
kinetic dominated bounces that give LQC corrections to the power spectra that are in 
the observational window.\footnote{Potential dominated bounces in quadratic
case give too much inflation and as a result the LQC corrections to the cosmological
perturbations do not fall in the observationally relevant window.} Therefore, tensor
modes--whose evolution equation does not depend on the potential--acquire 
the same LQC corrections for all $k$ for both potentials. 
The evolution equation of the scalar modes, on the other, includes an effective potential 
$\tilde{\mathcal U}$ (\eref{eq:scalar}) that involves the scalar field potential 
$V(\phi)$. Thus, despite the background evolution being similar for both potentials for 
kinetic dominated bounces, one might expect differences for the tensor perturbations. 
Nonetheless, it turns out that the magnitude of $\tilde{\mathcal U}$ for both 
potentials is too small to leave any imprint on the tensor power spectrum in the 
{\it observational window}. Hence, the observable LQC corrections to both the tensor and
scalar modes remain phenomenologically robust under the change of potential from 
the quadratic to the Starobinsky potential. Specifically, we found that for small 
$k$ modes, the scalar and tensor power spectra deviate from the standard Bunch-Davies 
ones, while for large $k$ modes they agree remarkably well. One can exploit the freedom
in choice of the initial state for the cosmological perturbations to match the 
$\sim 3 \sigma$ anomaly of power suppression observed in the CMB \cite{ag2}. Thus,
the power suppression anomaly observed in the CMB might have a quantum gravity origin.

\item {\bf Imprint on super horizon modes:} Although the effect of
$\tilde{\mathcal U}$ is too small to be in the observational window for both 
potentials, it can leave non-trivial imprints on the 
super horizon modes whose physical wavelengths are larger than the observable 
universe today. As shown in \fref{fig:rlqcbd}, the effect of $\tilde{\mathcal U}$
remains negligible for the quadratic potential while there are non-trivial
corrections for the Starobinsky potential. These features that are unique to the 
Starobinsky potential, could be important to study
the effects of non-Gaussian modulation \cite{schmidthui,agulloassym} and tensor 
fossils \cite{Dai:2013kra} on the power spectrum, where one considers coupling 
between observable and super horizon modes.

\end{itemize}

Note that the framework of cosmological perturbations on the quantum modified 
background used relies on the validity of the test field approximation. 
We have studied only those initial conditions that respect this approximation.
This excludes, however, a very small region of the initial data surface. 
It is possible that a generalization of the framework beyond the test field 
approximation might include potentially interesting cases which are excluded 
in our analysis. 

We will conclude with a few remarks and possible future extensions of the work
presented here. First, LQC provides a possible mechanism to explain the $\sim 3 \sigma$ 
anomaly observed in the CMB for low $k$, where the power in the tensor and scalar 
perturbations is lower than expected. Currently, the significance of this effect 
is limited by cosmic variance. However, the statistical significance of this LQC 
correction is expected to improve when taking into account the cross-correlation 
with the polarization data, which is expected to be released in the near future. 
This issue is currently being investigated and will be reported in future publications.
A second closely related avenue for future work is to address whether other anomalies
observed in the CMB could be explained by the same mechanism. This is a challenging 
and interesting problem. 

Third, we have followed the framework of quantum fields on quantum cosmological 
spacetimes according to which the dynamics of the background
spacetime relevant for perturbations can be encoded in the dressed metric given
in \eref{eq:dressed}. For computational simplifications, in this paper, we
assumed that the background spacetime is given by sharply peaked
wavefunctions for which the dressed metric can be well approximated by the
effective description of LQC. In other words, we have ignored the fluctuations
in the background geometry. It is important to note that the effective
description of LQC has only been derived for sharply peaked Gaussian states 
which become semi-classical at late times \cite{vt}. Non-trivial differences 
between the effective dynamics and the full LQC evolution arise 
for widely spread and non-Gaussian states \cite{dgs2,dgms}. Therefore, it is
natural to expect some corrections to the results obtained in this paper for
states which violate the assumptions in the derivation of the
effective description. There could be two leading sources of such corrections:
(i) corrections to the effective metric itself, which will require a more
generalized notion of the effective LQC description as obtained in \cite{ag1}, and
(ii) corrections due to approximating the dressed metric with the effective metric.
Both these issues have been investigated for the quadratic potential via
numerical simulations in \cite{aag1}. Fortunately, these results show 
that considering more generalized states for the background does not add 
new phenomenological parameters \cite{aag1}. The effect of fluctuations in the background 
states is degenerate with the initial condition for $\phi$ at the bounce. 
Specifically, the effect of fluctuations in the background state can be reproduced 
by adjusting the parameter $\phib$. Hence, the results are phenomenologically robust 
for more general background states. 
We expect a similar result for the Starobinsky potential, but this needs to be checked
explicitly. A full treatment of this problem requires evolution of 
wavefunctions, which is numerically challenging and left for future work. 

Moreover, particle physics issues have not been addressed in this work.
In particular, questions about the physical origin of the scalar field and how the
standard model of particle physics is created during reheating remain open.

Lastly, there is a mathematical equivalence between inflation with a Starobinsky 
potential and a modified theory of gravity. Specifically, via a conformal transformation,
one can write the action studied here (that is, the Einstein-Hilbert 
action plus a scalar field with a Starobinsky potential) as $\int \d^4 x \sqrt{-g} (R+
\f{1}{6M^2} R^2)$ \cite{staro,DeFelice:2010aj}. This mathematical equivalence is true 
\emph{classically} and there is no reason to believe that the quantization of each is 
equivalent too. For instance, see \cite{Kamenshchik} where it was shown that at first 
loop order the equivalence is no longer true off-shell. 
This paper is the first step towards extracting quantum gravitational implications of 
the Starobinsky model of inflation by studying the scalar field in a 
Starobsinky potential (known as `the Einstein frame'), 
where loop quantization is well understood. 
A complete treatment of the problem requires loop quantization of the modified 
gravitational theory as well (known as the `Jordan frame'). This is a difficult problem, 
but there are already interesting ideas to tackle this problem in LQC 
\cite{Zhang:2011vi,Zhang:2011qq}. Further analytical and 
numerical work is needed to understand the evolution of cosmological perturbations 
in the Jordan frame. The hope is that the main results of this paper will be similar
to those derived from quantizing the Starobinsky model in the Jordan frame.

\acknowledgements{We are grateful to Abhay Ashtekar for his guidance 
and feedback on this manuscript. We also would like to thank Ivan 
Agullo, Aurelien Barrau, Mairi Sakellariadou and Parampreet Singh for fruitful 
discussions, Suddhasattwa Brahma for discussions during the early stages of 
this work and John Barrow for correspondence. 
This research was partially supported by NSF grant PHY-1505411, the 
Eberly research funds of Penn State and a Frymoyer Fellowship to BB.}

%\bibliographystyle{kp}
%\bibliography{references/reference}
%\bibliography{references/lqc}

\end{document}